\documentclass[twocolumn]{aastex631}

\usepackage{amsmath,bm} 
\usepackage{xcolor} 

\begin{document}

\title{Unified weak lensing constraints on the evolution of the mass -- X-ray luminosity \\relation for galaxy clusters}

\author{I. Pederneiras}\email{isabelpederneiras@usp.br}
\affiliation{Universidade de São Paulo, Instituto de Astronomia, Geofísica e Ciências Atmosféricas, Rua do Matão, 1226, 05508-090 São Paulo, SP, Brazil}

\author{A. Finoguenov}
\affiliation{Department of Physics, University of Helsinki, P.O. Box 64, FI-00014 Helsinki, Finland}


\author{E. Cypriano}
\affiliation{Universidade de São Paulo, Instituto de Astronomia, Geofísica e Ciências Atmosféricas, Rua do Matão, 1226, 05508-090 São Paulo, SP, Brazil}

\author{J. Comparat}
\affiliation{Max-Planck-Institut für extraterrestrische Physik, Giessenbachstrasse 1, 85748 Garching, Germany}

\author{K. Kiiveri}
\affiliation{Department of Physics, University of Helsinki, P.O. Box 64, FI-00014 Helsinki, Finland}

\author{L. van Waerbeke}
\affiliation{Department of Physics and Astronomy, University of British Columbia, 6224 Agricultural Road, Vancouver, BC V6T 1Z1, Canada}

\author{R. Dupke}
\affiliation{Observatório Nacional, Rua General José Cristino, 77, São Cristóvão, 20921-400 Rio de Janeiro, RJ, Brazil}
\affiliation{Department of Astronomy, University of Michigan, 311 West Hall, 1085 South University Ave., Ann Arbor, MI 48109-1107}
\affiliation{Eureka Scientific, 2452 Delmer St. Suite 100, Oakland, CA 94602, USA}



\begin{abstract}
Scaling relations between galaxy cluster properties are crucial for understanding cosmology and baryonic physics. Rigorous calibration of the $M-L_X$ relation, employing weak lensing mass and consistent statistical methodology, is challenging due to heterogeneous cluster samples. The release of LEGACY imaging data introduced the possibility of unifying the cluster selection. We present the all-sky extension of the CODEX catalog based on LEGACY data and introduce a Bayesian framework for calibrating the X-ray luminosity–mass relation, derived for 100 clusters with weak lensing mass measurements. Using the X-ray luminosity estimates for those clusters from ROSAT All-Sky Survey (RASS) data, we perform a power-law fit to the $M-L_X$ relation. Furthermore, taking advantage of the recently released eROSITA data (eRASS1), we assess the impact of point source contamination on cluster fluxes for 42 clusters in the eRASS1 footprint. The RASS fit yields a slope of $\beta = 0.75 \pm 0.09$, 1.7$\sigma$ lower than the best self-similar prediction, with marginal evidence for the redshift evolution of the normalization ($\gamma = 0.65 \pm 0.43$). As for the eRASS1 analysis, the slope is substantially steeper, $\beta = 1.11 \pm 0.15$, and in further agreement with the prediction of self-similarity. No additional evolution is also seen ($\gamma = 0.004 \pm 0.790 $). While our results provide the practical means for cosmological studies of both RASS and eRASS data, the link to cluster physics is much cleaner after the cluster flux contamination is reduced. We also analyzed the impact of the selection function on calibration, finding that its full modeling is essential.

\end{abstract}



\section{Introduction} \label{sec:intro}

The current understanding of our expanding Universe suggests that primordial fluctuations, in an otherwise extremely homogeneous early Universe, inflated rapidly, intensifying density contrasts. Over time, gravitational instabilities in overdense regions led to the formation of large filamentary structures, where the collapse of dark matter halos through accretion occurs in the ’nodes’ of such filaments. In this scenario, the first structures formed would have been small and progressively merged during the evolution of the Universe \citep{PS74, White&Rees78, Blumenthal84}, hence placing galaxy clusters in the unique position as the most massive systems with a certain degree of dynamic maturity. 

Because galaxy clusters probe the history of structure formation in the Universe, they are often used to constrain cosmological parameters. More specifically, this can be achieved through the number density of clusters per halo mass and redshift, particularly sensitive to the $\Omega_M$ and $S_8$ parameters \citep{Allen11}. However, this approach requires obtaining the mass estimates for large samples of galaxy clusters, making it useful to adopt a scaling relation between the cluster mass and a more accessible observable.

An important feature of clusters is their hot, rarified gas of the intercluster medium (ICM), responsible for diffuse X-ray emission, almost entirely thermal \emph{bremsstrahlung}. In fact, since the 1980s, various X-ray surveys have been used to detect and investigate these objects. Cluster X-ray luminosity is known to scale with the cluster mass and is a common choice for obtaining tight cosmological constraints \cite[for a review see][]{CF}. The functional form of the $M-L_X$ relation is well-established by the self-similar model \citep{Kaiser86}, which is expected to follow a power law with exponent $\beta=4/3$. In reality, the picture is more complicated, as the thermodynamic properties of the clusters are affected by the feedback due to star formation and AGN activity as well as the diversity of cluster mass profiles \citep[e.g.,][]{Nagai06, Pratt09, Planelles15, Braspenning23}.

When studying mass-observable relations, it is essential to have a reliable mass calibration. As such, the weak gravitational lensing method offers great prospects for cluster mass estimates, since it accounts for the entire structure mass without relying on assumptions about the cluster’s dynamical state. Various works have calibrated the $M-L_X$ relation using hydrostatic mass estimates through X-ray or Sunyaev–Zel’dovich observations \citep[e.g.,][]{Chen07, Eckmiller11, Bulbul19, Lovisari20}, or using simulations \citep{Braspenning23}. In contrast, works based on weak gravitational lensing masses are not as common \citep[e.g.,][]{Lea10, Kett15}, primarily due to the need for deep, high-resolution images and accurate photometric redshifts. Furthermore, the $M-L_X$ redshift evolution can benefit from further studies, with the currently published results being quite uncertain \citep[e.g.,][]{Lea10, Bulbul19}.

For a meaningful study of scaling relations, one should seek to analyze high-quality data, with unbiased methods to estimate the observable quantities. On the other hand, to improve precision, it is preferable to work with large samples, with parameters spanning an extensive value range \citep{Dawson19, Witte17}. Unfortunately, these two aspects are difficult to reconcile in most astronomical studies. To access these requirements, we take advantage of new developments in cluster identification, enabled by LEGACY surveys (LS), to unify the selection of clusters for lensing studies. The deepest all-sky X-ray source catalog currently available to us comes from the ROSAT All-Sky Survey, which we use as a base for our study. 
We present an extension of the CODEX catalog to cover most of the extragalactic sky, using the redMaPPer run on LEGACY DR9 (LS9) and DR10 (LS10) data. 
To calibrate the $M-L_X$ relation, we use three previously published samples with reliable weak lensing mass estimates while using the CODEX cluster catalog to compute the sampling function. We developed a hierarchical Bayesian model for our calibration, which is an ideal approach for describing a complete model of the data, accounting for systematic and measurement errors. Furthermore, we analyze the impact of some statistical assumptions and factors in the Bayesian model on the final calibration.

This paper is structured as follows: in section 2, we present the CODEX-LS cluster catalog and describe our weak lensing sample; in section 3, we discuss the current state of research and previous studies on the $M-L_X$ relation; our Bayesian statistical model is detailed in section 4; section 5 is dedicated to our results and discussions, where we also compare with other calibrations in the literature; finally, we conclude and summarize in section 6. Throughout this work, $\log$ denotes the logarithm in base 10. We assume a flat $\Lambda$CDM cosmology with $\Omega_m=0.3$, $ \Omega_{\Lambda}=0.7$ and $H_0 = 70\;\mathrm{km\;s}^{-1}\;\mathrm{Mpc}^{-1}$, unless stated otherwise. 

\section{Data}\label{samples}

\subsection{The CODEX-LS catalog}
The COnstrain Dark Energy X-ray (CODEX) catalog performs X-ray selection of galaxy clusters based on the third Data Release of all-sky ROSAT survey data \citep[RASS DR3][]{Voges99} with a flux limit reaching $10^{-13}\;\mathrm{erg\;s}^{-1}\;\mathrm{cm}^{-2}$ in the 0.5--2.0 keV band \citep{Finoguenov2020}. 
For the catalog construction, the X-ray source list was obtained by finding detections above the $4\sigma$ detection significance level, using the \emph{wavelet} image decomposition \citep{Vikh98} on spatial scales up to $6^\prime$, which are suitable for cluster characterization at the redshift of the study. Larger spatial scales were employed for low-z group studies in the AXES project \citep{Khalil24,Damsted24}. 
To identify X-ray sources as clusters, version 8 (Python) of the red-sequence Matched-filter Probabilistic Percolation cluster-finding algorithm code \citep[redMaPPer,][]{Ry14} is run in scan mode \citep[e.g., as in][]{ider2020cosmological,Kluge2024}, utilizing the photometric data from the 9th (Northern Hemisphere) and 10th Data Release of the DECam Legacy Survey \citep[DECaLS,][]{dey2019overview}. 
To construct a clean cluster catalog, we have applied the cluster selection criteria outlined in \citet{Finoguenov2020} (reproduced in Eq.\ref{Pclean} below) and used the efforts to unify the determination of cluster richness across different datasets included in the LS data release. This was done for sources North of Declination $+32^\circ$ (DR9 north), where the richness values were rescaled by 2\% to account for the systematic difference in the filters.

The CODEX-LS cluster redshifts span from 0.0 to 1, but for our choice of X-ray catalog, the practical range is $0.1<z<0.7$ \citep{Finoguenov2020}. The low redshift cut is also supported by \citet{Damsted23}, which finds a significant increase of the scatter in the X-ray luminosity for redshifts below this threshold. We apply the cluster selection outlined in Eq.\ref{Pclean}, along with mask fraction (MF$<0.3$) and redshift ($z<0.7$) cuts, to clean the catalog and avoid contamination, with further details presented in section \ref{Selections}.

The CODEX-LS catalog covers 14 thousand square degrees and has over 4 thousand sources in the clean catalog.

\subsection{Weak lensing sample}

\subsubsection{The main-subsample}
The main sample adopted by this work was based on \citet{Kiiveri21}, which selected a high-richness and high-z subsample of CODEX galaxy clusters with follow-up observations with the Wide Field Optical Imaging MegaCam as part of the CFHT Legacy Survey \citep{Cibirka17}. The work included additional CODEX clusters identified in the CFHTLS observed fields, covering a larger range of redshift and richness. Among all these clusters, 35 present weak lensing mass estimations computed using the pipeline developed by \citet{Kiiveri21} and \citet{Gruen15}. This pipeline was made available for this work, providing access to the details regarding the computation of the mass estimates for this subsample and enabling us to use the complete modeling of the mass probability distribution function (PDF).

A simple cosmology correction was made to their work since they adopt a mass density value of $\Omega_m = 0.27$, which differs from our choice of $0.3$. Their mass estimation analysis includes a complex dependence on the mass density parameter. However, we found it sufficient to consider the general weak lensing formalism, where the $\Omega_m$ dependence relies on the estimated critical surface density \citep[for more on this see the review of ][]{NARAYAN95}. Hence, when changing the $\Omega_m$ value, the mass estimation would be rescaled by the ratio of the mass densities. As the primary uncertainty arises from the number of background galaxies, this correction does not significantly affect the results. Nevertheless, we include this cosmology change in the systematic error for consistency.

When considering the cleaned CODEX catalog (described in section \ref{Selections}), we ended up with a CODEX-CFHT weak lensing subsample of 27 clusters. Hereafter, we refer to it as the \emph{main-subsample}. The sample also extends to higher redshift ranges, up to 0.65. This is a distinction between previous works and an important feature since we attempt to investigate the $M-L_X$ redshift evolution.

\subsubsection{Additional weak lensing samples}

\citet{Herbonnet20} computed the weak lensing mass of 100 X-ray selected clusters from the Multi-Epoch Nearby Cluster Survey (MENeaCS) and the Canadian Cluster Comparison Project \citep[CCCP,][]{Hoek15}, using CFHT images. From those, 83 clusters are present in the CODEX catalog, and 46 are selected after applying the cuts to clean the catalog. This final subsample - hereinafter \emph{Herbonnet20} - comprises massive clusters (over $\sim 3.2 \cdot 10^{14}M_{\odot}$), reaching redshifts values of $z \sim 0.55$.

The hundreds of shear-selected clusters from \citet{Oguri21} are also considered in our work. They used the Hyper Suprime-Cam Subaru Strategic Program (S19A) to compute the clusters' mass via weak gravitational lensing. Although these clusters are not X-ray selected, the statistical analysis can be adapted to CODEX (discussed in Section \ref{Selections}) since their catalog selection is deeper than that of RASS. From their work, 84 clusters are also part of the extended CODEX catalog, and a final sample of 27 clusters are present in the cleaned catalog - hereinafter \emph{Oguri21}. These are also high mass clusters (over $\sim 2.1 \cdot 10^{14}M_{\odot}$) and span a redshift range up to $0.58$.

All cluster matching between these works and the CODEX catalog was made considering a 3-arcminute distance from the clusters' center and a redshift interval of $\Delta z = 0.05$. In summary, our final weak lensing (WL) sample consists of 100 CODEX galaxy clusters with weak lensing mass estimates and X-ray information. Basic data of these clusters is available in Tables \ref{tab:catalogue_columns} and \ref{tab:catalogue2_columns}.

\begin{figure}[h!]
    \centering
    \includegraphics[width=\hsize]{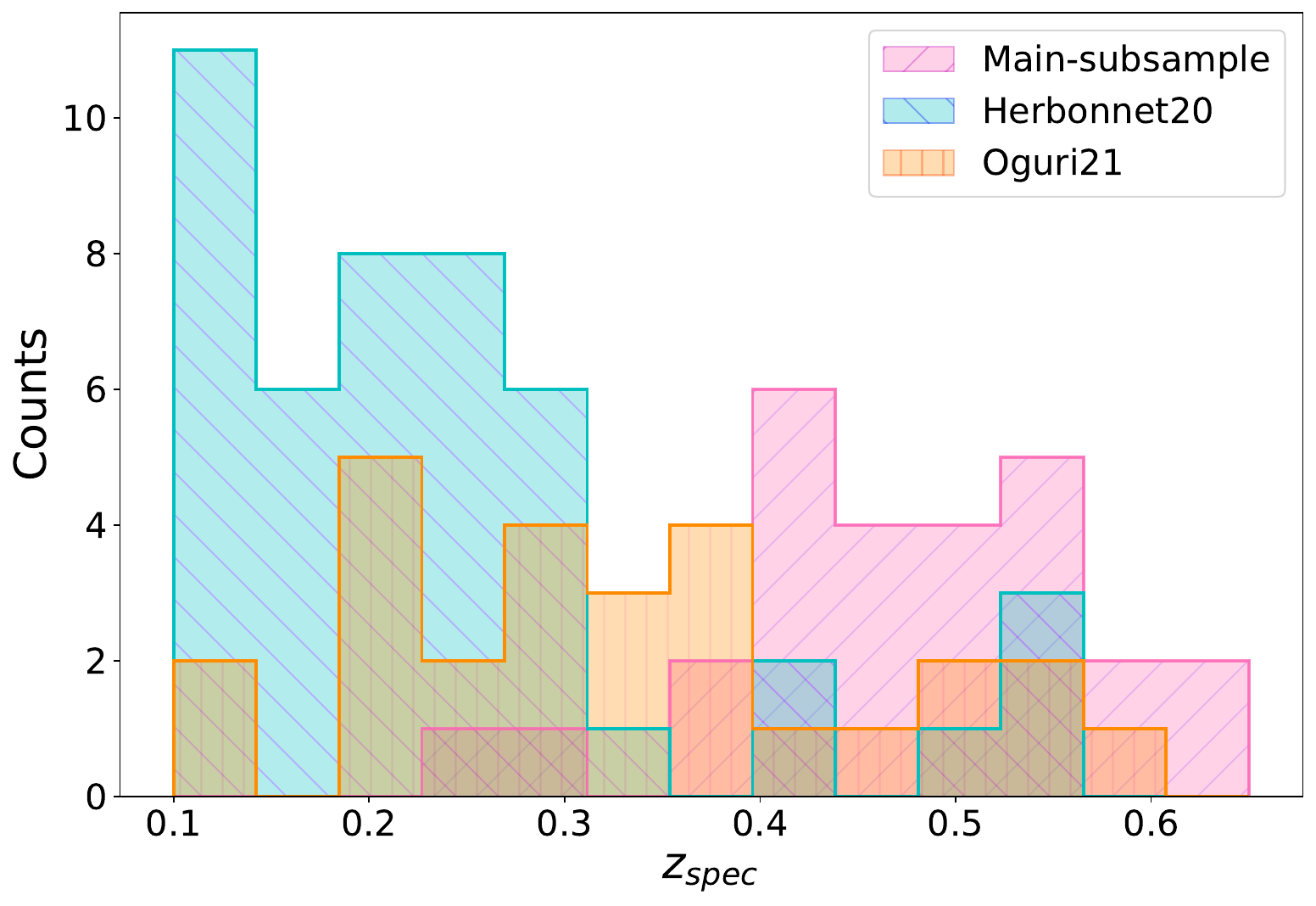}
    \caption{Spectroscopic redshift distribution of the \emph{WL sample}. The colours blue, orange and pink represents the \emph{Herbonnet20}, \emph{Oguri21}, and \emph{main-subsample}, respectively.}
    \label{hist_z}
\end{figure}

We also show in Fig. \ref{hist_z} the redshift distribution of each subsample, where we observe that the \emph{main-subsample} introduced by this work considerably increases the number of high-redshift galaxy clusters.

\section{The $M - L_X$ scaling relation}\label{MlxScaling}

Due to the complexity of galaxy cluster systems, their formation is often modeled using non-dissipative approaches that effectively predict correlations between cluster properties. One key model is the \emph{self-similar} model proposed by \citet{Kaiser86}, which assumes that clusters form through gravitational collapse with density fluctuations following a power-law dependence on their size, and no additional physical processes introducing scale dependence. Under this framework, clusters are considered scaled versions of one another, leading to the expectation that the main cluster properties (e.g., $L_X$, $T_X$, velocity dispersion) follow power laws.

To make the $M - L_X$ relation explicit, apart from self-similarity we must make other assumptions such as hydrostatic equilibrium and that the X-ray luminosity arises from thermal \emph{bremsstrahlung} emission \citep[for more on this, see the review from][]{Kravt12}. As a power law, the scaling relation is more commonly worked in logarithmic space as a linear relation. Such correlation is described in Eq. \ref{LxM}, where we also added the parameter $\gamma$ to account for the redshift evolution: 
\begin{equation}\label{LxM}
    \log\left[\frac{L_X E(z)^{-1}}{L_{X,\mathrm{piv}}}\right]=\alpha + \beta\:\log\left[\frac{M E(z)}{M_{\mathrm{piv}}}\right] + \gamma \:\log\left[\frac{(1+z)}{(1+z_{\mathrm{piv}})}\right].
\end{equation}\\
For this work, we chose the pivot values of $L_{X,\mathrm{piv}} = 10^{44.64}\;\mathrm{erg s}^{-1}$, $M_{\mathrm{piv}} =10^{14.84} M_{\odot}$, and $z_{\mathrm{piv}} = 0.29$, which corresponds to the median values of  X-ray luminosity, mass, and redshift of the \emph{WL sample} (all 100 clusters).
We also note that the term $E(z) = H(z)/H_0$ dictates the dynamics of the Universe and is a natural consequence of its expansion. Hence, it is not responsible for describing the evolution of a scaling relation, even though it depends on the redshift.

Considering Eq. \ref{LxM} for X-ray luminosities in soft energy bands\footnote{When using soft-band X-ray data, comparisons with self-similarity must avoid the $\beta = 4/3$ prediction, which applies only to bolometric luminosity.}, where the emissivity is almost independent of temperature, self-similarity predicts $\beta=1$ and no additional redshift evolution - i.e., $\gamma=0$. Moreover, \citet{Lovisari21} reported the impact of the cluster's temperature ($T_X$) and metallicity ($Z$) in flattening this expected slope. According to their work, when considering typical values for massive systems (e.g., $Z = 0.3Z_{\odot}$, $T_X$ between $3-10$ keV) and observing in the $0.1-2.4$ keV band, the self-similar prediction should be $\beta_{\mathrm{self}} = 0.9$.

Nonetheless, the purely gravitational model by Kaiser assumes temperature and density profiles as independent of mass, which is not rigorously observed \citep{NFW97, Nagai06, Ascasibar06}. Contrary to the self-similar theory, when accounting for non-gravitational contributions to the ICM, deviations from equilibrium are expected to be mass-dependent, and so are the cluster's thermodynamic profiles. \citet{Braspenning23} used hydrodynamic FLAMINGO simulations to analyze X-ray properties, revealing a strong mass dependence of these profiles. While not their primary focus, they also examined the $M - L_X$ relation and found a mild difference in slope between low and high-mass regions, with a steeper slope in the galaxy group regime.
Furthermore, X-ray observations also indicate a possible increase in the gas mass fraction with total mass, such that the lower mass systems would present a smaller amount of gas, and consequently a lower X-ray emission \citep{Pratt09, Planelles15}. This observational trend could result in the steepening of the $M - L_X$ slope, especially when analyzing the galaxy group regime.
At last, the calibration focuses on studying clusters through a simple two-observable relation ($M$ and $L_X$) despite evidence of hidden parameters. \citet{Fujita19} emphasized the concentration parameter's role in flattening the $M-L_X$ relation and increasing the luminosity scatter, potentially impacting results within limited mass ranges.

Various works have calibrated the $M - L_X$ relation using hydrostatic mass estimates (i.e., through X-ray or Sunyaev–Zel’dovich observations) and the consensus seems to be a steeper value for the slope when compared with the self-similar prediction, both when analyzing low-mass \citep{Chen07,Eckmiller11} and high-mass \citep{Bulbul19,Lovisari20} systems. Although less common, works utilizing weak gravitational lensing masses to calibrate the $M - L_X$ relation have also found the slope to be steeper than self-similarity, whether studying galaxy groups \citep{Lea10} or a combination of groups and clusters \citep{Kett15}.

\section{Hierarchical model for $M-L_X$ calibration}\label{Hier_model}

The scaling relation between the X-ray luminosity and total mass of galaxy clusters, as well as its redshift evolution, is given by Eq. \ref{LxM}, providing a relation between our observable $l_X=\log L_X$ and the true mass $\mu=\log M$, given a model with parameters $\bm{\theta}= \{\alpha,\beta,\gamma,\sigma_{\mathrm{intr}}\}$. The joint probability distribution that there is a cluster of total mass $\mu$, at a specific redshift $z$, can be described as \citep{Kelly07}:

\begin{equation}\label{P(s,mu|theta,z)}
    \begin{aligned}
    P(l_X,\mu|\bm{\theta},z)=P(l_X|\mu,\bm{\theta})P(\mu|z)P(z).
    \end{aligned}
\end{equation}\\
In the above equation, the term $P(\mu|z)$ is the halo mass function (HMF), providing the number density of dark matter halos (e.g., of galaxy clusters) as a function of mass:

\begin{equation}\label{HMF}
    \frac{dn}{dM} = f(\sigma)\frac{\bar{\rho}}{M}\frac{d\ln\sigma^{-1}}{d\ln M}\;,
\end{equation}

\noindent where $\sigma$ is the \emph{rms} variance, $\bar{\rho}$ is the mean matter density, and $f(\sigma)$ is the so called halo multiplicity function. In this work, we adopt the HMF calibrated by \citet{Tinker08}. 

The term $P(z)$ in Eq. \ref{P(s,mu|theta,z)} describes the differential comoving volume, defined as

\begin{equation}
    \begin{aligned}
    &\frac{dV(z)}{dz}= \frac{c}{H_0}\frac{D^2(z)}{E(z)}\;;
    \end{aligned}
\end{equation}\\
where
\begin{equation}
    \begin{aligned}
        D(z) = \frac{c}{H_0}\int_0^z \frac{dz}{E(z)}\;\;\; \text{is the comoving distance.}
    \end{aligned}
\end{equation}
Both terms combined account for the conditional probability of the existence of a galaxy cluster with a given mass and redshift.

As for the term $P(l_X|\mu,\bm{\theta})$, it provides the probability of having a scattered X-ray luminosity $l_X$ given an underlying true luminosity - i.e., expected luminosity obtained by equation \ref{LxM} given a mass value and model parameters, $\langle l_X|\mu,z,\bm{\theta}\rangle$. It is reasonable to assume that the data is normally scattered around its theoretical expected value \citep{Sereno16, Allen11}. Thus, our function is modeled as a log-normal distribution centered on the expected value $\langle l_X|\mu,z,\bm{\theta}\rangle$ with the Gaussian form for deviations characterized by the intrinsic scatter parameter, $\sigma_{\mathrm{intr}}$.

\begin{equation}
    \begin{aligned}
        &P(l_X|\mu,z,\bm{\theta}) = \frac{1}{\sqrt{2\pi} \sigma_{\mathrm{intr}}}\exp \left[-\frac{1}{2}\frac{(l_X - \langle l_X|\mu,z,\bm{\theta}\rangle)^2}{\sigma_{\mathrm{intr}}^2}\right];
    \end{aligned}
\end{equation}\\
where
\begin{equation}
    \begin{aligned}
        \langle l_X|\mu,z,\bm{\theta}\rangle =\, &\alpha +\beta\:\log\left[\frac{M E(z)}{M_{\mathrm{piv}}}\right] + \gamma \:\log\left[\frac{(1+z)}{(1+z_{\mathrm{piv}})}\right] \\
        &+\log[E(z)L_{X,\mathrm{piv}}].
    \end{aligned}
\end{equation}\\
When effectively incorporating all terms in Eq. \ref{P(s,mu|theta,z)}, we are also addressing significant selection biases (i.e., Malmquist and Eddington biases). 
Another important consideration is the effect caused by measurement errors since we cannot directly access the observable quantity. Thus, we must model the probability distribution of measured observables (denoted by $\Tilde{l_X},\Tilde{\mu},\Tilde{z}$) given scattered observables ($l_X,\mu,z$).
For our case, it is reasonable to consider the cluster's spectroscopic redshift equal to its true redshift. As such, $P(\Tilde{z}|z)$ is modeled as a Dirac delta function and will not be explicit in the following equations. 
The probability of measuring a cluster mass $\Tilde{\mu}$ through gravitational lensing is given by the mass PDF $P(\Tilde{\mu}|\mu)$. For the \emph{main-subsample} we follow the mass estimation pipeline done in \citet{Kiiveri21}. They assumed the weak lensing mass likelihood as proportional to the probability of measuring a mean surface mass density contrast $\Delta\Sigma$ for a cluster of true mass $M_{\mathrm{200c}}$. According to their work, the mass probability density function can be expressed by a multivariate Gaussian distribution:

\begin{equation}\label{MPDF}
    \begin{aligned}
        P(\Delta\Sigma|M) \propto \frac{1}{\sqrt{\det C}}\exp\left(-\frac{1}{2} E(M)^{T}C^{-1}E(M)\right)\;,
    \end{aligned}
\end{equation}
where
\begin{equation}
    \begin{aligned}
        E(M) = \Delta\Sigma_{\mathrm{obs}} -\Delta\Sigma_{\mathrm{model}}\;,
    \end{aligned}
\end{equation}
and where $C$ is a covariance matrix accounting for several physical effects that can lead to uncertainties (e.g., galaxy shape noise, halo triaxiality, correlated secondary halos, and off-centering). We then use Eq. \ref{MPDF} in the logarithmic space for all clusters in the  \emph{main-subsample}, defining $\mu = \log M_{\mathrm{200c}}$.

\begin{figure}[h!]
    \centering
    \includegraphics[width=\hsize]{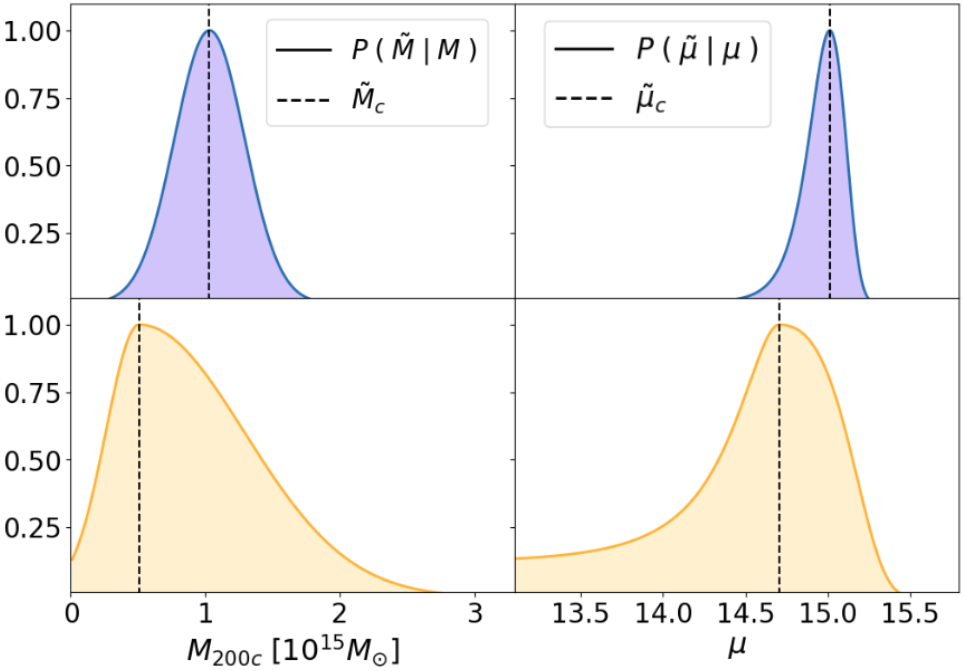}
    \caption{The mass probability distribution function for a cluster from the \emph{Herbonnet20} subsample (\emph{upper panels}) and from the \emph{Oguri21} subsample (\emph{lower panels}). The \emph{left hand} panels are in the linear space, while the \emph{right hand} panels are in the logarithmic space.}
    \label{mpdf_OH}
\end{figure}

For the other two subsamples, \emph{Herbonnet20} and \emph{Oguri21}, we use the information available in the respective catalogs and model the mass PDFs as Gaussian functions in the linear space, $P(\Tilde{M_c}|M_c)$. These normal distributions are centered at the estimated weak lensing mass values with deviations given by the measurement errors. Specifically for \emph{Oguri21} subsample, because the errors are not symmetric in either linear or logarithmic space, we model them as a combination of two Gaussians. Fig. \ref{mpdf_OH} illustrates the normal functions described above for clusters of the \emph{Herbonnet20} and \emph{Oguri21} subsamples, as well as the respective mass PDFs in log space $P(\Tilde{\mu}|\mu)$. 

As for the X-ray luminosity distribution, a fairly common approach is to model it as a normal distribution in linear space. Although this is reasonable at low redshifts, where the photon counts are abundant, it becomes an inaccurate approach when extending to higher redshifts. At such a regime, a Poisson distribution is a more accurate representation given the small number of photon counts. Our cluster sample ranges from $z = 0.1$ to $z = 0.65$, and we also want to avoid any possible bias on the redshift evolution of the scaling relation. With that in mind, we adopt the probability of observing the source counts ($\Tilde{\eta}$) given the expected number count ($\eta$) as a Poisson distribution:

\begin{equation}\label{eta}
    \begin{aligned}
        P(\Tilde{\eta}|\eta)=\frac{(\eta)^{\Tilde{\eta}}e^{-\eta}}{\Tilde{\eta}!}\,,
    \end{aligned}
\end{equation}

\noindent where the photon counts $\eta$ is given by
\begin{equation}
    \begin{aligned}
        \eta (L_X)=\frac{L_X}{4\pi D_L^2K(T,z)} \cdot \frac{T_{\mathrm{exp}}}{C(nH)}.
    \end{aligned}
\end{equation}\\
In the expression above, $D_L$ is the luminosity distance, $K(T,z)$ is the K correction, $T_{\mathrm{exp}}$ is the exposure time, and $C(nH)$ is the flux conversion due to absorption by the H column in the line of sight. All photon count estimation also accounts for background and aperture corrections. Hence, Eq. \ref{eta} describes, in effect, the term $P(\Tilde{l_X}|l_X)$.

With these considerations, we modify Eq. \ref{P(s,mu|theta,z)} and marginalize it over all true quantities, obtaining

\begin{equation}\label{P(til s,mu,z|theta)}
    \begin{aligned}
        P(\Tilde{l_X},\Tilde{\mu},\Tilde{z}|\bm{\theta})=&\\
        \int dl_Xd\mu&\underbrace{P(\Tilde{\eta}|\eta)\,P(\Tilde{\mu}|\mu)P(l_X|\mu,\bm{\theta})}_{\substack{\text{Accounts for measurement}\\\text{errors and intrinsic scatter}}}\underbrace{P(\mu|\Tilde{z})}_{\text{HMF}}\underbrace{P(\Tilde{z})}_{\text{$\frac{dV}{dz}$}}.
    \end{aligned}
\end{equation}

\subsection{Selection effects}

Apart from the considerations above, we must also contemplate the effects of the selections made during the construction of our subsample. In a general context, we can introduce the variable $I$ as a condition for whether or not the cluster passed the selection. As such, the probability of having the measured observables $\bm{\Tilde{o}}=\{\Tilde{l_X},\Tilde{\mu},\Tilde{z}\}$ given that the cluster passed the selection is

\begin{equation}\label{P(Tilde o|I,theta)}
    \begin{aligned}
        P(\bm{\Tilde{o}}|I,\bm{\theta})=\frac{P(I|\bm{\Tilde{o}},\bm{\theta})\,P(\bm{\Tilde{o}}|\bm{\theta})}{P(I|\bm{\theta})}.
    \end{aligned}
\end{equation}\\
In the end, what we want is the probability distribution of parameters given the observables $P(\bm{\theta}|\bm{\Tilde{o}})$. According to Bayes theorem, Eq. \ref{P(Tilde o|I,theta)} is actually the likelihood function $L(\bm{\Tilde{o}}|I,\bm{\theta})$ we need to model:
\begin{equation}\label{P(theta| til o)}
    \begin{aligned}
        P(\bm{\theta}|\bm{\Tilde{o}}) \propto \; \pi(\bm{\theta})\; L(\bm{\Tilde{o}}|I,\bm{\theta}),
    \end{aligned}
\end{equation}
where $\pi(\bm{\theta})$ are the priors for the parameters, which are assumed to be independent.

We can see that the term $P(\bm{\Tilde{o}}|\bm{\theta})$ in Eq. \ref{P(Tilde o|I,theta)} represents the probability of obtaining the set of observables $\bm{\Tilde{o}}$ given our model. This is the expression we already defined in Eq. \ref{P(til s,mu,z|theta)}. Also, the general probability of all clusters passing the selection - denoted by $P(I|\bm{\theta})$ - can be described by simply marginalizing the numerator over all observables.

Finally, we can write our full likelihood function as:

\begin{equation}\label{L}
    \begin{aligned}
        L(\Tilde{l}_X,\Tilde{\mu},\Tilde{z}|I,\bm{\theta})\propto\;&\int dl_X\,d\mu\,\; P(I|\bm{\Tilde{o}},\bm{\theta})\\
         &\cdot P(\Tilde{\eta}|\eta)\, P(\Tilde{\mu}|\mu)\\
           &\cdot P(l_X|\mu,\bm{\theta})P(\mu|\Tilde{z})P(\Tilde{z}).
    \end{aligned}
\end{equation}\\
Which depends on the combined sample selection, now represented by the general term $P(I|\bm{\Tilde{o}},\bm{\theta})$, but which will be unraveled in the following section.

\subsubsection{The selection functions}\label{Selections}

In this section, we give an overview of the selection functions used in the construction of our cluster subsample, some of which are further detailed in \citet{Finoguenov2020}. 
First, two richness selections were applied to the CODEX catalog sample. In \citet{Finoguenov2020}, they simulated the completeness limits for the CODEX catalog using RASS and SDSS data in each 0.1-width bin of redshift. It was found that the $10\%$ completeness limit for RASS matches the \citet{Klein19} definition of low contamination sample - a $5\%$ contamination when identifying RASS sources using the Dark Energy Survey (DES).
To account for the $10\%$ completeness, we must then remove the sources for which the observed richness $\Tilde{\lambda}$ is below this limit. We achieved this by applying the following selection:

\begin{equation}\label{Pclean}
    \begin{aligned}
        P(I_{\mathrm{RASS}}|\Tilde{\lambda},z)=
        \begin{cases}
            1, & \text{if}\ \Tilde{\lambda}>22\left(\frac{z}{0.15}\right)^{0.8} \\
            0, & \text{otherwise}
        \end{cases}\;.
    \end{aligned}
\end{equation}\\
In addition, their work found that for $z < 0.2$ and $z>0.5$, it was important to account for the completeness of the optical cluster finder. To model this selection function, we follow the analytical form adopted in \citet{Finoguenov2020}, obtained based on the tabulation of \citet{Ry14}. We suppress the redshift dependence of their function since we are using deeper optical data, and the sensitivity of high-z cluster detection is driven by the RASS data \citep{ider2020cosmological}, so the probability of redMaPPer cluster detection is
\begin{equation}\label{Popt}
    \begin{aligned}
        P(I_{\mathrm{opt}}|\ln\,\lambda)=1-\frac{1}{2}\,\text{erfc}\left(\frac{\ln\lambda -\ln\lambda_{50\%}}{\sqrt{2}\,\sigma}\right),
    \end{aligned}
\end{equation}

\noindent where $\ln\lambda_{50\%}=\ln(17.2)$ and $\sigma = 0.2$ for the CODEX catalog.

In establishing the weak lensing sampling function of clusters, we remove the clusters with uncertain richness estimates. The origin of high uncertainty is associated with a high cluster mask fraction (MF) due to the presence of bright stars, and so the corresponding selection is MF $ < 0.3$.

\begin{figure}[h!]
    \centering
    \includegraphics[width=\hsize]{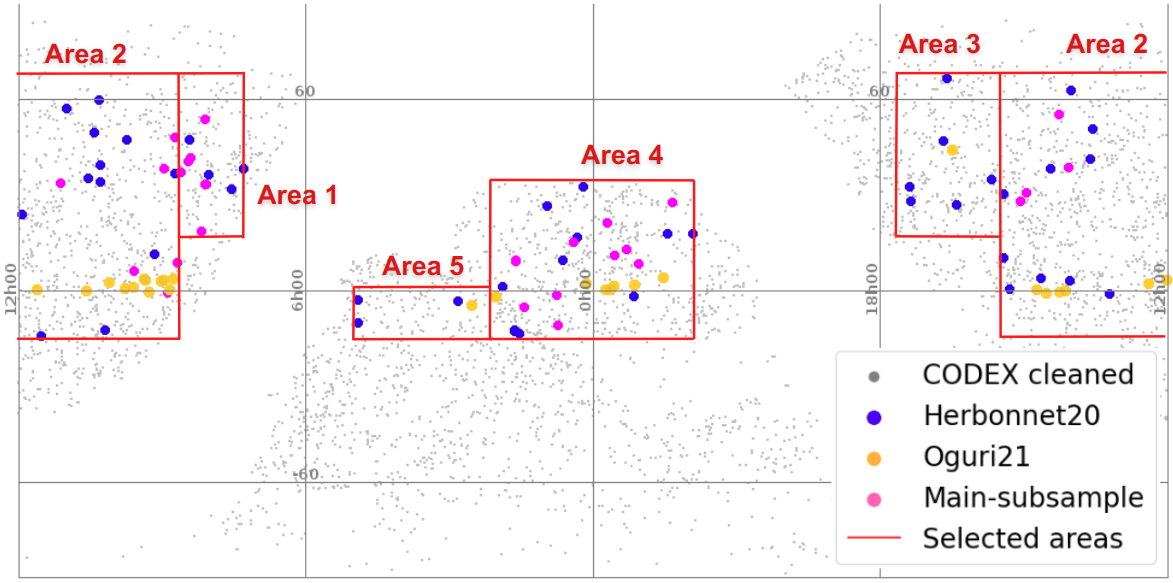}
    \caption{Projection of CODEX samples' coordinates onto a Cartesian plane, along with the delineation of the five specified areas in red. The pink, blue, and orange dots represent the \emph{main-subsample}, the \emph{Herbonnet20}, and the \emph{Oguri21} subsamples, respectively. The gray dots represent the other galaxy clusters in the cleaned CODEX catalog.}
    \label{clean_areas}
\end{figure}

Furthermore, for our sample to be considered representative of the cluster population, the ratio between the number of our \emph{WL sample} and the total clusters in the cleaned CODEX catalog should be $\sim 1$. However, this is far from the truth. To mitigate this issue, we defined a smaller survey area for the analysis, concentrated in the sky regions where our weak lensing subsamples lie. Figure \ref{clean_areas} shows the Cartesian projection of the samples' coordinates and the five areas defined for this purpose; the final survey area used in this work is simply the combination of all of them.

With these considerations, we can now implement a subsample selection function. We neglect all clusters with richness below 60 to reduce the effect of low-richness clusters on the scaling relation, which would only be present at low but not at high redshifts. Additionally, to define $\Tilde{\lambda}\geq 60$, we use a ratio of the \emph{WL sample} to the total number of the clean CODEX clusters inside the defined survey area. We bin the samples by richness into equal logarithmic bin widths and compute the ratio of the height of the bins. After, we fit a linear piecewise function between the mean of each bin, obtaining:

\begin{equation}\label{Psamp}
    \begin{aligned}
    P(I_{\mathrm{samp}}|\Tilde{\lambda})=
    \begin{cases}
        0, & \Tilde{\lambda}<60 \\
        \frac{3.9}{1000}\Tilde{\lambda}-\frac{201}{1000}, & 60\leq \Tilde{\lambda}<143.5 \\
        \frac{0.1}{1000}\Tilde{\lambda}+\frac{205}{1000}, & 143.5\leq \Tilde{\lambda}<258.6 \\
        \frac{500}{1000}, & 258.6\leq \Tilde{\lambda}<342
    \end{cases}
    \end{aligned}
\end{equation}

\begin{figure}
    \centering
    \includegraphics[width=\hsize]{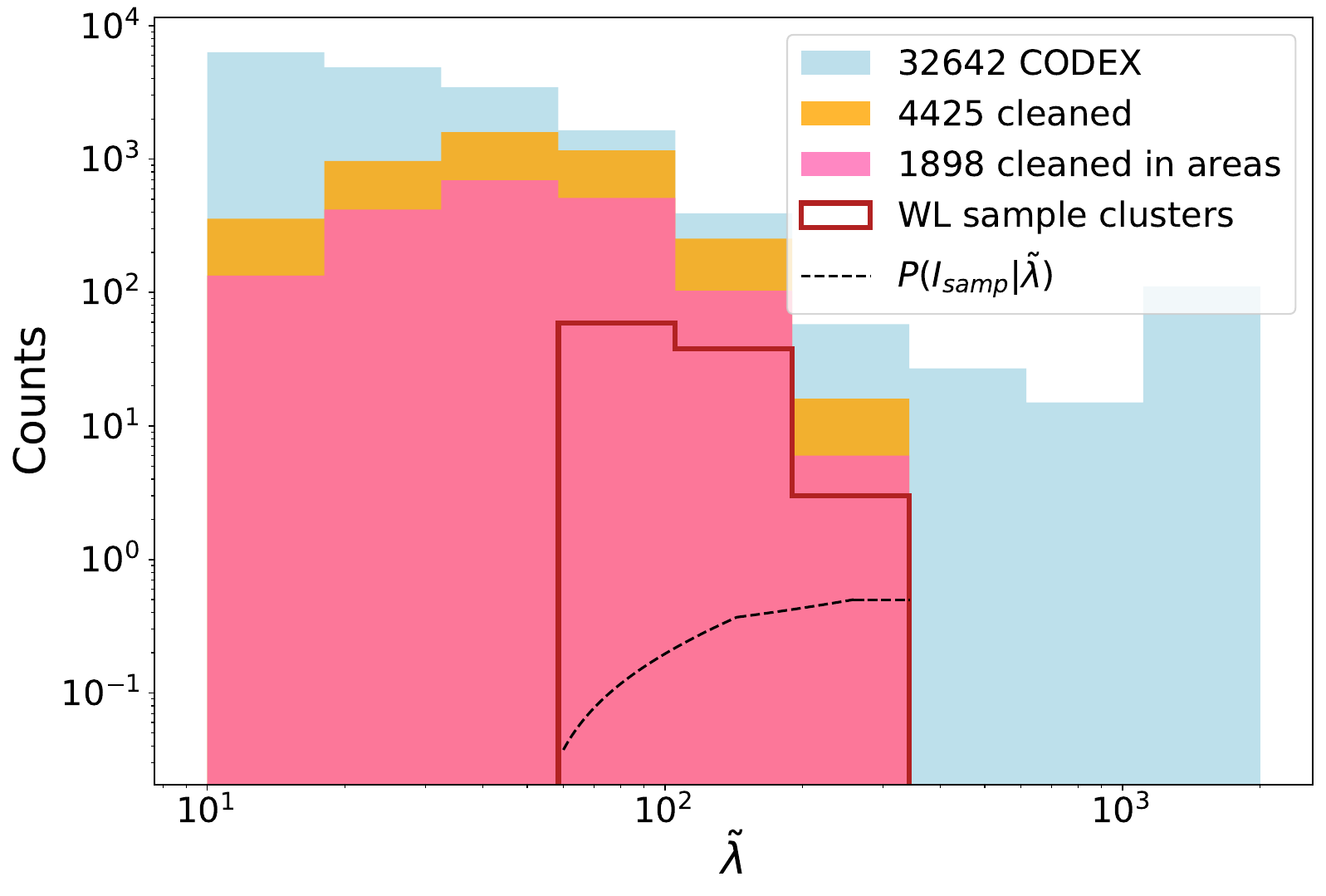}
    \caption{Richness distribution of the CODEX sample in bins of equal width. The main difference between the cleaned and the total sample is the removal of clusters with richness affected by the poor quality of optical data. The area sampling returns a very similar richness distribution. The weak lensing cluster sampling function is represented by the dashed black line.}
    \label{codex_sampling}
\end{figure}

Figure \ref{codex_sampling} illustrates this selection.

Finally, there is also the X-ray selection function, associated with the depth of RASS data and the source detection in the CODEX survey, for which we give a brief overview \citep[for more on this, see ][]{Finoguenov2020}.

This CODEX selection function returns the effective survey area for the catalog. Given certain observables, we use it to correct the probability of detecting a galaxy cluster in a comoving volume. This selection considers the probability of detecting a source given the observed source counts - $\eta^{ob}$ - and the shape parameters of a \hbox{$\beta$-profile} surface brightness distribution - $\beta(\mu)$ and $r_c=\ln\,R_c$ \footnote{A $\beta$-profile surface brightness distribution with radius $r$ is given by $S_B=S_0\left[1+\left(\frac{r}{R_c}\right)^2\right]^{-3\beta+0.5}$}. 

The function also accounts for a correlation between the X-ray luminosity and the cluster's richness, and for the cluster shape being covariant to the scatter in the $M-L_X$ relation. The probability of observing source counts given the expected number count is considered as well and is modeled similarly to Eq. \ref{eta}.

In effect, this CODEX X-ray selection function is available as a grid, resulting from simulations that account for the probabilities described above. The function returns the effective survey area and the parameter values used to calculate it. In the end, interpolating this simulated grid can give us a function for the probability of detecting a galaxy cluster with luminosity $l_X$, redshift $z$, and intrinsic scatter $\sigma_{\mathrm{intr}}$ in the CODEX catalog, denoted by $P(I_X|\Tilde{z},l_X,\sigma_{\mathrm{intr}})$.

At last, we can write the full selection functions as follows:

\begin{equation}\label{Select.}
    \begin{aligned}
        P(I|\Tilde{\lambda}, \Tilde{z}, \sigma_{\mathrm{intr}})=&\int\, d\mu\,dl_X\,d\lambda\\
        &\cdot P(I_{\mathrm{RASS}}|\Tilde{\lambda},\Tilde{z}) \xleftarrow{\text{CODEX identification}} \\
        &\cdot P(I_{\mathrm{opt}}|\Tilde{\lambda}) \xleftarrow{\text{completeness of redMaPPer}} \\
        &\cdot P(I_{\mathrm{samp}}|\Tilde{\lambda}) \xleftarrow{\text{weak lensing sampling}}\\
        &\cdot P(I_X|\Tilde{z},l_X,\sigma_{\mathrm{intr}})\xleftarrow{\text{CODEX selection function}}\\[2mm]
        &\cdot P(\Tilde{\lambda}|\lambda)\,P(\lambda | \langle \ln\lambda | \mu \rangle)\;.
    \end{aligned}
\end{equation}\\

Because some of the selections are richness-dependent, we introduced the functions $P(\Tilde{\lambda}|\lambda)$ and $P(\lambda | \langle \ln\lambda | \mu \rangle)$ to account for the richness measurement errors and intrinsic scatter, respectively. Following the same reasoning discussed in the previous section, the former expression is modeled as a normal distribution to account for the redMaPPer measured error $\delta_{\lambda}$. As for the latter, we adopt a log-normal distribution for the richness given its expected value $\langle \ln\lambda | \mu \rangle$:

\begin{equation}\label{rich_loggauss}
    \begin{aligned}
        &P(\lambda | \langle \ln\lambda | \mu \rangle) = \frac{1}{\sqrt{2\pi} \lambda \sigma_{\lambda_{\mathrm{intr}}}}\text{exp} \left[-\frac{1}{2}\frac{(\langle \ln\lambda|\mu\rangle - \ln\lambda)^2}{\sigma_{\lambda_{\mathrm{intr}}}^2}\right]\;.
    \end{aligned}
\end{equation}\\
The above expression is obtained using the $\lambda - M$ relation calibrated by \citet{Kiiveri21} for clusters in the CODEX catalog. According to their work, $\sigma_{\lambda_{\mathrm{intr}}} = 0.17$ and $\langle \ln\lambda | \mu \rangle$ is computed from

\begin{equation}\label{lamb_m}
    \ln(\lambda) = 4.42 + 0.49\;\ln\left(\frac{M}{10^{14.81}M_{\odot}}\right).
\end{equation}\\

We assessed the impact of adopting the richness–mass relation from literature by exploring the uncertainty range of $\sigma_{\lambda_{\mathrm{intr}}}$, which captures deviations from the relation’s parameters (see Appendix \ref{richmass} for details). This only mildly affects the normalization $\alpha$, with negligible impact on the slope and evolution parameters.

\subsection{Priors}\label{priors}

In our analysis, we adopt \emph{flat uninformative priors} for the intercept parameter, and the mass and redshift slopes - i.e., a constant value for

\begin{centering}
    $-24 < \alpha < 10$,\\
    $0 < \beta < 2.5$,\\
    $-5 < \gamma < 5$.\\
\end{centering}

As for the intrinsic scatter parameter $\sigma_{\mathrm{intr}}$, there seems to be no overall agreement about the choices for weakly informative priors in the literature. Even so, the Jeffreys prior is one of the most common approaches. With that in mind, we adopted the \emph{modified Jeffreys prior} \citep{Gregory}, which avoids the divergence at small $\sigma$ found in the original function:

\begin{equation}\label{priorsig_m}
    \pi(\sigma) = \frac{1}{(\sigma +a)\ln[(\sigma_{\mathrm{max}}+a)/a]}\;,
\end{equation}
where we defined $a=0.03$, for which the function's mean value is at the somewhat expected value of $\sigma_{\mathrm{intr}}=0.33$ \citep{Lea10}. We also tested our entire analysis using the $\chi^2$, gamma, and inverse gamma distributions as uninformative priors for the scatter parameter but found no significant differences in the results.

In conclusion, we model three likelihood functions ($L_i$) for each of our subsamples (i.e., \emph{main-subsample}, \emph{Herbonnet20}, and \emph{Oguri21}). These are given by the general equation \ref{L} with the discussed selections in Eq. \ref{Select.}. At last, the posterior distribution we need to maximize to estimate our best parameters is, in log space,

\begin{equation}
    \begin{aligned}
        \log P(\bm{\theta}|\bm{\Tilde{o}}) \propto \log \pi(\bm{\theta}) + {\sum_i}\log L_{i}.
    \end{aligned}
\end{equation}

\section{Results}\label{results}
To sample our posterior distribution for the parameters, we used the Python package \textsc{emcee} \citep{GoodmanWeare10, ForMac13}, which consists of a Monte Carlo Markov Chain algorithm. We found that it was enough to execute 24 chains with 1,500 steps each and burn the first 350 steps that it takes for the chains to converge. The best values and their errors were computed as the median and standard deviation of the distributions for each marginalized parameter, respectively. 

\begin{figure}[h!]
    \centering
    \includegraphics[width=\hsize]{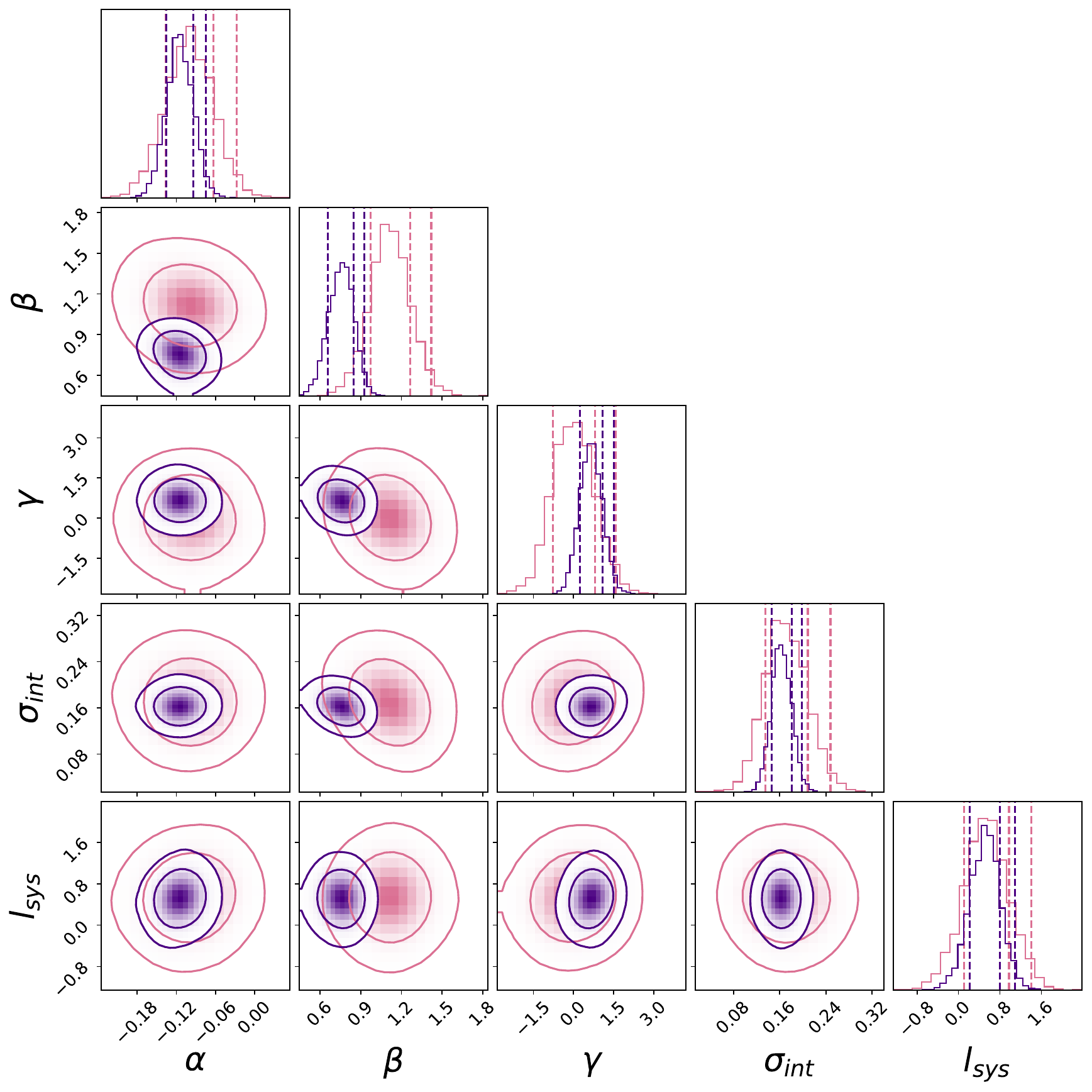}
    \caption{Best parameters' values from the MCMC fitting for the main fit (\emph{purple}) and eRASS1 fit (\emph{pink}). The one and two-dimensional projections of the posterior distributions are also shown. The contours represent the $1\sigma$ (68\%) and $2\sigma$ (95\%) confidence regions.}
    \label{best_params}
\end{figure}

We applied our general Bayesian analysis for the 100 clusters in our \emph{WL sample} with the uninformative priors discussed in section \ref{priors}, which will be hereafter referred to as the main fit. Our results are displayed in Table \ref{table_results1} and Fig. \ref{best_params} shows the best parameters and their correlation in a triangle plot. Apart from our model parameters, we also report the best-fit value for the lensing systematic uncertainty $l_{\mathrm{sys}}$, which serves as a control parameter from the weak lensing mass estimation of the \emph{main-subsample}, as defined in \citet{Kiiveri21}.

\begin{table}[h!]
\begin{center}
 \caption{Parameters, their initial values, priors, and posteriors from the MCMC fitting with all samples.}
 \label{table_results1}
 \begin{tabular}{lcccc}
  \hline
  \hline
  Parameter & Initial & Prior & Posterior\\
  \hline
  Intercept $\alpha$    &    -1.29  & flat(-24, 10) & $-0.12 \pm 0.02$ \\
  Slope $\beta$    &    0.9  & flat(0, 3) & $0.75 \pm 0.09$ \\
  Evolution $\gamma$    &    0  & flat(-5, 5) & $0.65 \pm 0.43$ \\
  Scatter $\sigma_{intr}$    &    0.33  & Jeff. prior & $0.16 \pm 0.02$ \\
  Systematic error $l_{sys}$    &    0  & N(0,1) & $0.47 \pm 0.31$ \\
  \hline
  \hline
 \end{tabular}
     
\end{center}
\end{table}

We found a slope value of $\beta = 0.75 \pm 0.09$, which is shallower than expected by self-similarity ($\sim 0.9$) with a difference of 1.7$\sigma$. The low value might be due to our focus on high-mass systems (Section \ref{samples}), which is discussed throughout these next sections.

The intrinsic scatter value was also smaller than that reported by most previous works - over $0.25$ \citep[e.g.,][]{Kett15, Lea10, Eckmiller11}. However, it is expected to have a smaller scatter when analyzing a high redshift sample \citep{Damsted23}. 

As for the redshift evolution parameter, we found a value of $\gamma = 0.65 \,\pm \,0.43$, which agrees with the no evolution prediction of zero at a $1.5\sigma$ level.

In essence, despite the modest slope value that might suggest the influence of unknown systematic factors, our findings do not render it statistically significant to sustain an evolution hypothesis of the $M-L_X$ relation. Using the MCMC chains, we also estimated that a positive value for the evolution parameter is preferred, with $P(\gamma > 0) = 94.37\%$.

\begin{figure}[h!]
    \centering
    \includegraphics[width=\hsize]{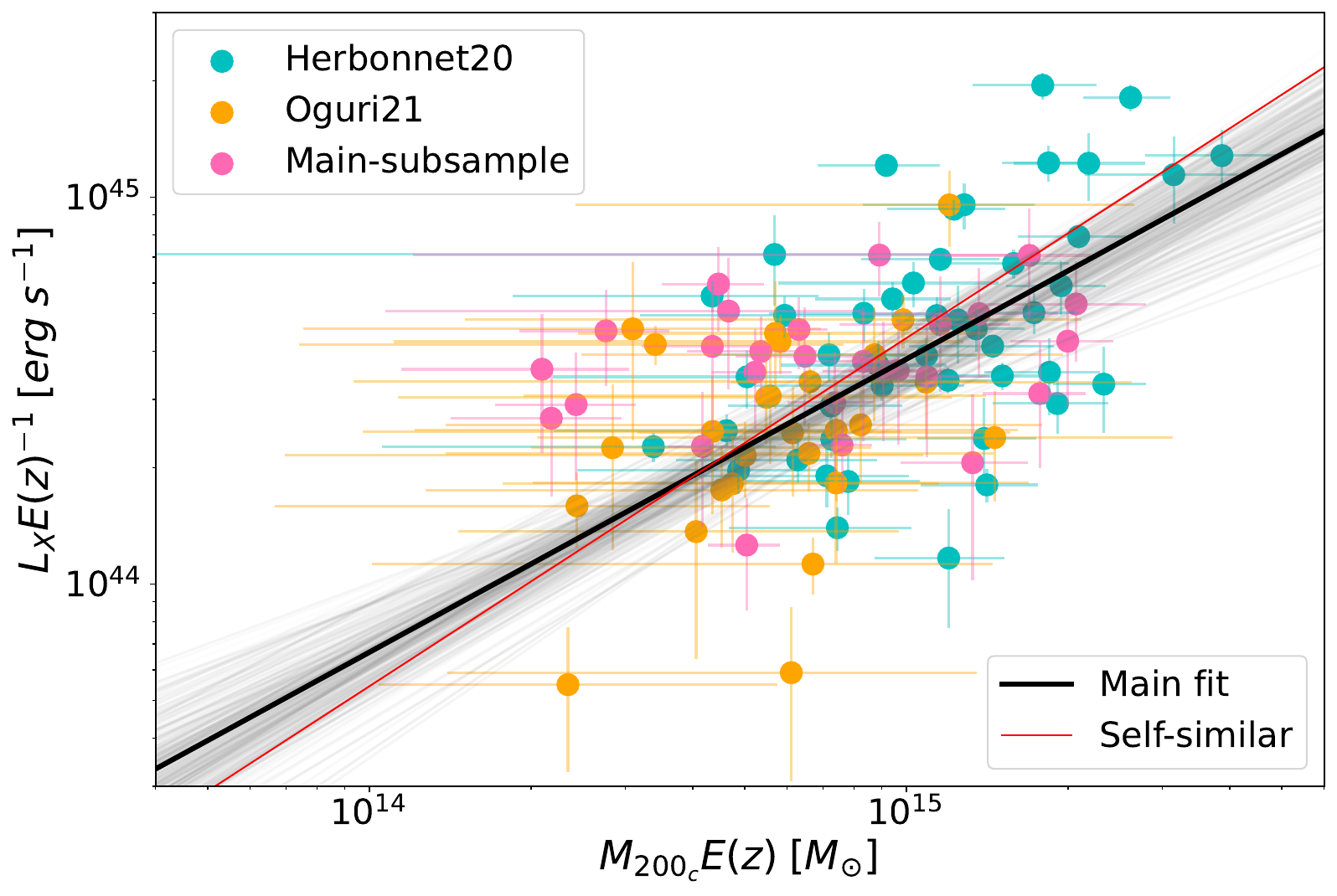}
    \label{subfig:a}
    \centering
    \includegraphics[width=\hsize]{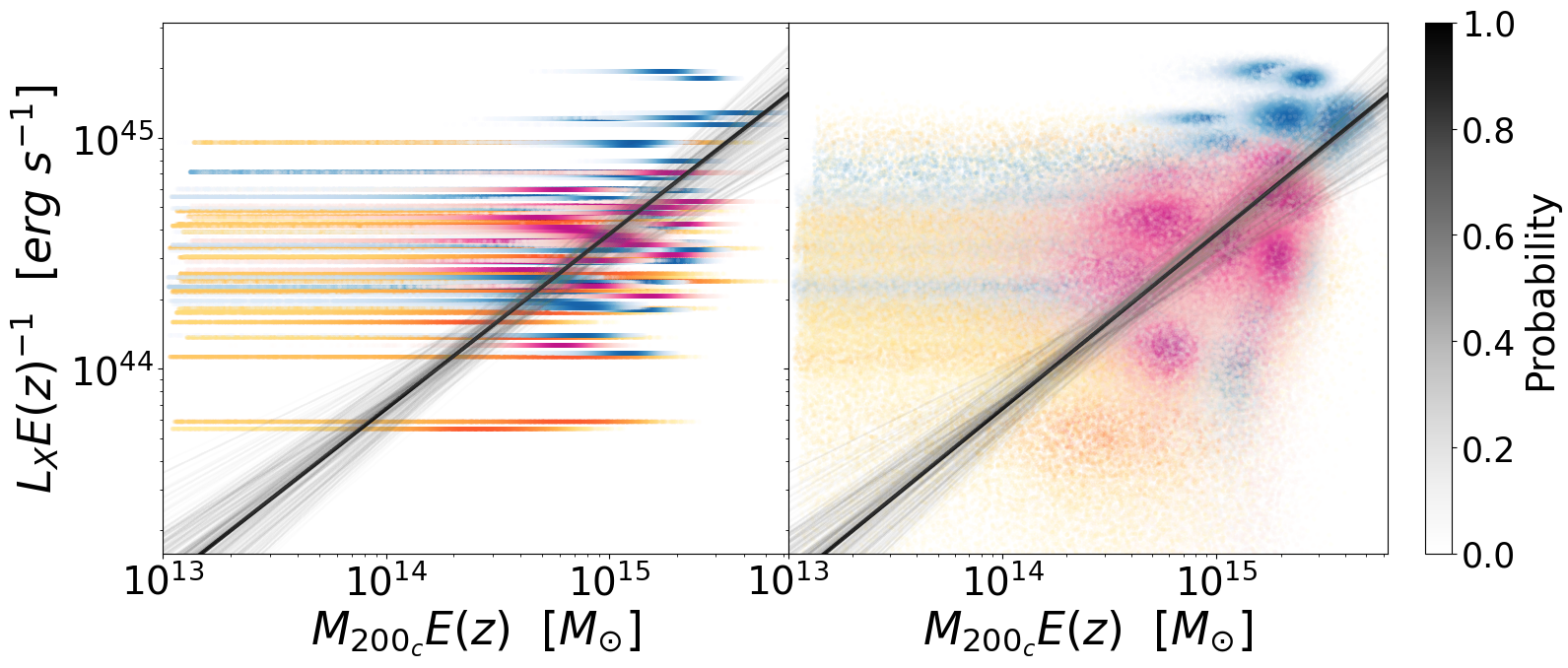}
    \label{subfig:b}
    \caption{The $M-L_X$ relation in log space. The black line represents the projected best fit obtained from the MCMC. The results from a few iterations are depicted by the faded gray lines, illustrating the error associated with the best fit. There is also a simple linear fit using the Python function \textsc{lmfit.minimize} with slope fixed at 0.9, which illustrates the self-similar prediction for comparison. The subsamples of \textit{Herbonnet20}, \textit{Oguri21}, and \textit{Main-subsample} are represented in blue, orange, and pink, respectively. In the \emph{lower-left} panel, only the mass probability distribution functions for each cluster are displayed. In the \emph{lower-right} panel, both the mass and X-ray luminosity PDFs are presented. The legends in the upper panel also apply to the lower panels.}
    \label{main_fit}
\end{figure}

In Fig. \ref{main_fit}, we show our best fit for the \emph{WL sample} projected on the logarithmic mass -- X-ray luminosity plane. The lower panels highlight the fact that we are using probability distribution functions in our analysis, by representing the mass PDFs (\emph{lower-left}) and both the mass and X-ray luminosity PDFs (\emph{lower-right}) for each cluster. The PDF representation might provide a clearer visualization of the relation between the best fit and the data.

To assess how the calibration is affected by our statistical approach, we also tested different features of our model (full analysis in Appendix \ref{elem_stat_model}). We concluded that incorporating selection functions leads to the steepening of the $M-L_X$ scaling relation. Our results suggest that, particularly for cluster samples selected based on richness, accounting for the richness-mass correlation plays a significant role. Additionally, we found evidence that the inclusion of the HMF helps constrain the evolution parameter $\gamma$. We also verified that its value does not change even when the slope is fixed to the self-similar prediction, $\beta_{\mathrm{self}} = 0.9$.

\subsection{Contribution of each subsample}\label{cont_samples}
We also aimed to verify the contribution of each subsample in the calibration of the $M-L_X$ relation. We would expect a significant impact of the \emph{main-subsample} in constraining the redshift evolution parameter $\gamma$ since its redshift range is complementary to the other subsamples. On the other hand, the larger measurement errors in the mass estimates of \emph{Oguri21} subsample could affect the constraints in the slope and intersection of the scaling relation. 

To examine these questions, we applied our Bayesian statistical model with uninformative priors to three different sets of cluster samples: All \emph{WL sample} but the \emph{main-subsample}; all but the \emph{Herbonnet20} subsample; and all but the \emph{Oguri21} subsample. We note that the only adjustment made to our model was on the sampling selection, Eq. \ref{Psamp}, where we kept the same approach, modifying only the sample groups accordingly.
The results for each of the three sets are listed in Table \ref{table_sum_all_fits}.

Contrary to what was expected, the inclusion of the \emph{main-subsample} has no major impact on constraining the evolution parameter $\gamma$. Additionally, when excluding the \emph{Oguri21} subsample the difference to the results from the main fit is under $1\sigma$. Upon analysis, it becomes evident that the \emph{Herbonnet20} subsample had the most impact on our results, where we found a difference in the evolution parameter of 2.4$\sigma$ - while the other parameters all agree under 1.5$\sigma$\footnote{Note that the intercept values in Table \ref{table_sum_all_fits} are not directly comparable since they are rescaled by pivot values, which depends on the slope and evolution parameters ($\alpha_{\mathrm{scaled}} = \alpha - \log L_{X,\mathrm{piv}} + \beta \log M_{\mathrm{piv}} + \gamma \log(1+z_{\mathrm{piv}})$).}. We conjecture that this might be due to a higher rate of cool core contamination in the \emph{Herbonnet20} subsample and further investigate this in Appendix \ref{H20}.

\subsection{Improving X-ray luminosity estimates using eROSITA data}

Our work utilizes the CODEX catalog, which extends the RASS flux limits to unprecedented values, and for which we can properly model selection functions to our analysis. Comparable flux limits have been achieved by the eROSITA (Extended Roentgen Survey with an Imaging Telescope Array) first data release \citep[eROSITA DR1, ][]{eRASS1}, at a significantly better spatial resolution, which presents us with an opportunity to improve X-ray flux estimates for our study of the $M-L_X$ relation.

The publicly released year 1 eROSITA-DE data (hereinafter eRASS1) cover only half of the sky and there is a small difference in the distribution of lensing data, leading to obtaining 42 matches (from the 100 \emph{WL sample}). We used the 0.6-2.3 keV band images of eROSITA, together with the officially released exposure and background files. We run {\it wvdetect} \citep{Vikhlinin09} to separate the unresolved emission from the cluster emission. The sources detected on spatial scales up to 32 arcseconds are excised from the flux extraction, a procedure similar to other published eROSITA analyses. As eROSITA luminosity, we report the flux extracted on the scales of wavelet detection of 4 arcminutes, performing the same flux extrapolation procedure as for the RASS sources. The changes introduced with eROSITA flux estimates consist of a substantially reduced contamination from the point sources, higher photon statistics, and improved characterization of cluster emitting zone, which can also lead to higher inferred luminosity due to the flux extrapolation procedure. In calculating the eROSITA luminosities, we produced a K-correction table, linking the countrates in the 0.6-2.3 keV band with the rest-frame 0.1-2.4 keV flux, and verified that our conversion is reproduced by the XSPEC luminosity calculation under the same cosmological assumptions. The general behavior of K-correction closely resembles the values in \citet{Jones_1998}, which is a standard reference for K-correction for X-ray clusters.
The matching of eRASS1 and RASS sources is done using the coordinates of X-ray centers, allowing for 3$^{\arcmin}$ uncertainty on RASS position.

\begin{figure}[h!]
    \centering
    \includegraphics[width=\hsize]{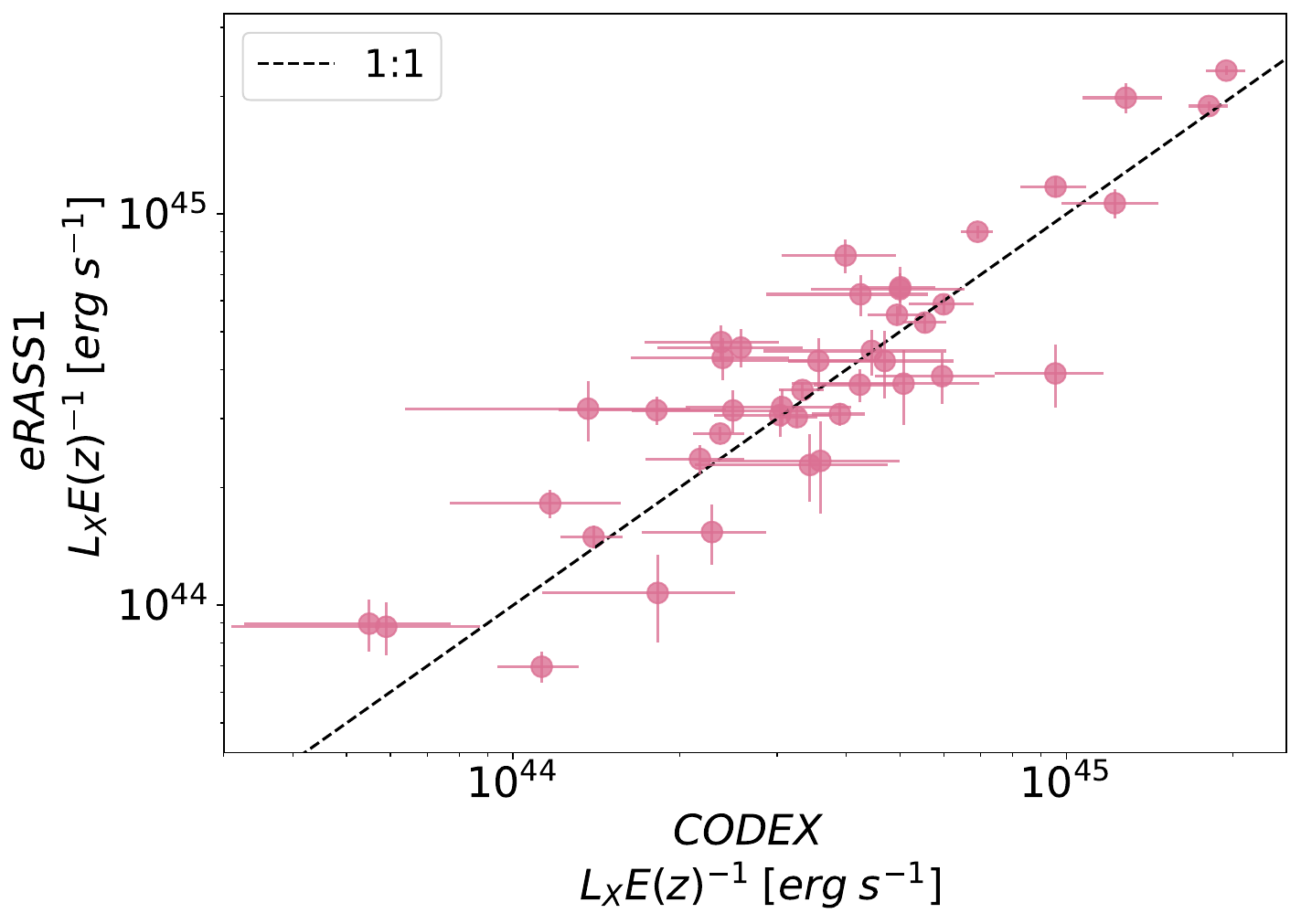}
    \caption{Comparison between eROSITA and CODEX X-ray luminosity estimates for a subsample of 42 galaxy clusters. The dashed curve describes the unity line.}
    \label{erositaXcodex}
\end{figure}

Figure \ref{erositaXcodex} compares the X-ray luminosities between CODEX and eROSITA for this \emph{eRASS1 sample} of 42 clusters. We find a general agreement between the two estimates, with a scatter of $0.16$ dex, and with eROSITA errors being significantly smaller — on average, these errors are a factor of two lower than those of CODEX.

We then proceeded to apply our $M-L_X$ calibration using eROSITA X-ray luminosities. Since the subsample is also CODEX selected, only two changes were made to our statistical model. We altered the sampling selection defined in Eq. \ref{Psamp} according to this new subsample. Additionally, we modeled the X-ray luminosity PDF as a normal distribution instead of a Poisson distribution, given eRASS1 statistics. As discussed in section \ref{rich_and_lxpdf}, this modification is expected to have minimal impact on the results.

Maintaining the uninformative priors outlined in section \ref{priors} and the overall MCMC setup, we performed the fit on the \emph{eRASS1 sample}. The results are displayed in Table \ref{table_sum_all_fits} and the posterior distributions are shown in Fig. \ref{best_params}. We found a slope value of $\beta = 1.11 \pm 0.15$, which concurs with our main fit and the self-similar prediction (within 2$\sigma$ and 1$\sigma$, respectively). The evolution parameter, $\gamma = 0.004 \pm 0.790$, is also consistent with our main fit finding of a non-evolution scenario.

\begin{figure}[h!]
    \centering
    \includegraphics[width=\hsize]{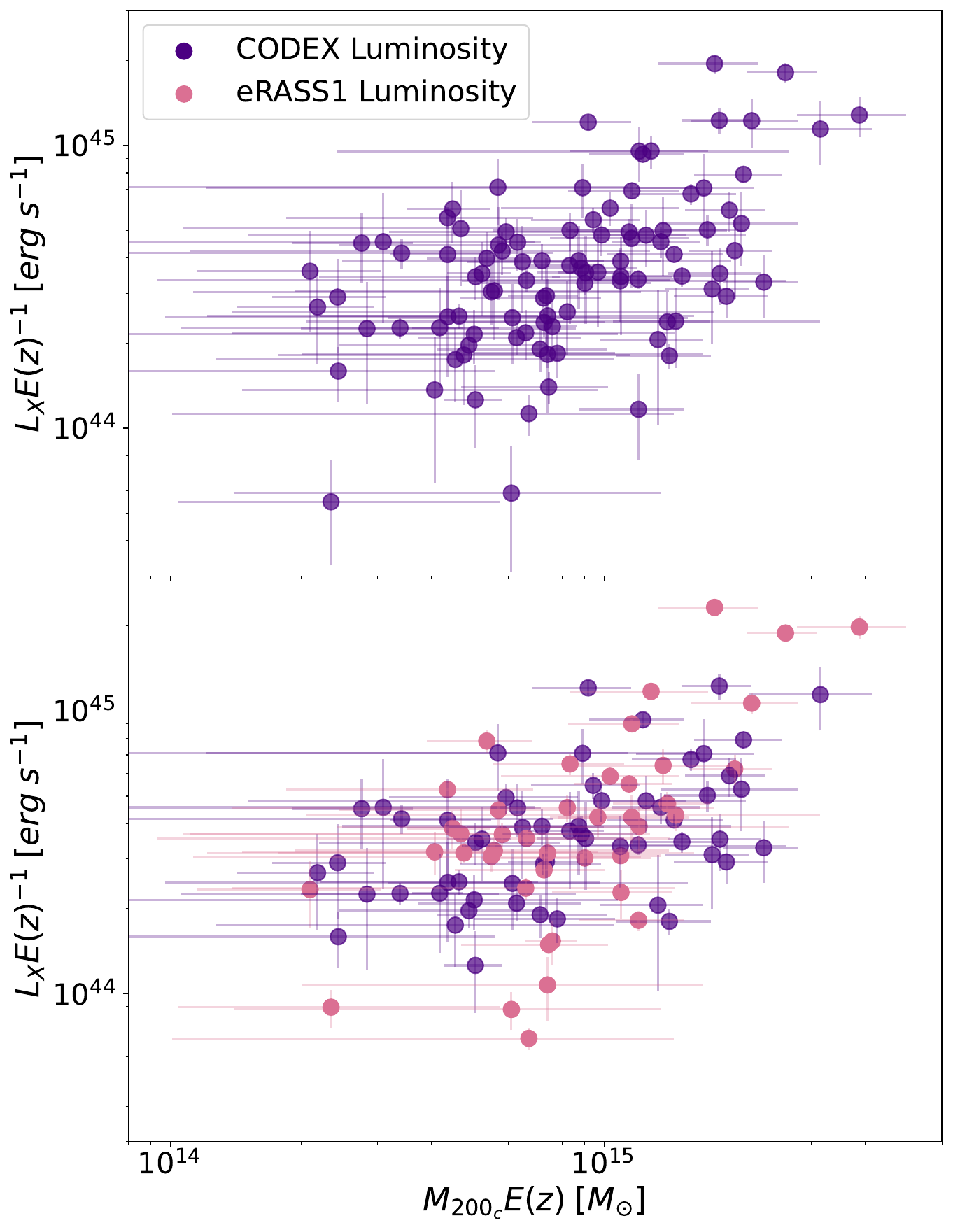}
    \caption{The \emph{WL sample} with CODEX (\emph{upper panel}) or CODEX and eRASS1 (\emph{lower panel}) X-ray luminosities in the $\log L_X - \log M$ plane. The clusters with CODEX $L_X$ estimations and those with eRASS1 $L_X$ estimations are represented in purple and pink circles, respectively.}
    \label{CODEX_eRASS1_M_Lx}
\end{figure}

Figure \ref{CODEX_eRASS1_M_Lx} displays the \emph{WL sample} and the matched eRASS1 clusters in the logarithmic $M-L_X$ plane to assess whether there might be some selection bias that justifies the increase of the slope value. Upon a visual analysis, no obvious source of bias could be observed. We conjecture that this change is likely due to flux contamination by point sources, which was accounted for in the eRASS1 $L_X$ estimates.

\subsection{Comparison with the literature}\label{comp_lit}

\begin{figure}[h!]
    \centering
    \includegraphics[width=\hsize]{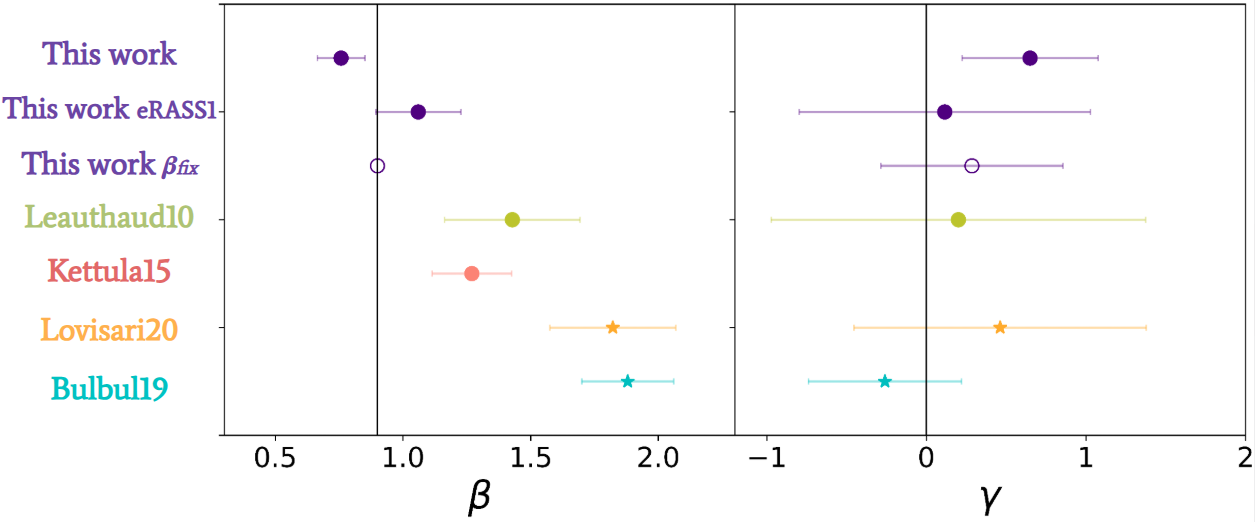}
    \caption{Values for the slope $\beta$ and evolution parameter $\gamma$ of the $M-L_X$ scaling relation. We compare the results obtained in this work (of the \emph{WL sample}, with and without a fixed slope value, and with the \emph{eRASS1 sample}) with some of the ones available in the literature - i.e., \citet{Lea10, Kett15, Lovisari20} and \citet{Bulbul19}. The vertical lines represent the self-similar prediction on the slope ($\beta = 0.9$) and the non-evolution expected value ($\gamma=0$). The star marker differentiates the works that used hydrostatic mass estimates in their analysis.}
    \label{beta_gamma_lit}
\end{figure}

It is important to note that our main fit result for the slope parameter is not consistent with previous works in the literature. However, the agreement improves when the comparison is made with the eRASS1 result. In Fig. \ref{beta_gamma_lit}, we compare our values for slope $\beta$ and the evolution parameter $\gamma$ with the ones obtained by \citet{Lea10} and \citet{Kett15}, who relied on weak lensing mass estimates, as well as with the outcomes of \citet{Lovisari20} and \citet{Bulbul19}, who employed hydrostatic mass values. These previous studies also utilized soft-band X-ray luminosities, most in the 0.1–2.4 keV range (as in our work), except for \citet{Bulbul19}, which used the 0.5–2 keV band. The difference between these bands has minimal impact on the expected self-similar slope value \citep[0.9 for the former and 0.93 for the latter, as discussed in][]{Lovisari21}, allowing for a direct comparison of their results with our work. We also note that \citet{Bulbul19} calibrated the $M-L_X$ relation using masses inferred by the Compton $Y$ measurements (i.e., integrated pressure), which could be reflected in the difference observed between our results. For the sake of coherence, our subsequent discussion focuses primarily on the two studies with weak lensing masses.

In \citet{Lea10}, the analysis was done using 206 galaxy groups (i.e., $M_{\mathrm{200c}} \lesssim 10^{14}M_{\odot}$) of the COSMOS catalog, where the gravitational lens mass estimate was performed through stacking in 9 bins of X-ray luminosity and redshift. Their work found a slope of $\beta = 1.42 \pm 0.18$, which is steeper compared with the self-similar prediction of $\sim 0.9$. Furthermore, their work also analyzed the redshift evolution of the $L_X-M$ relation in this group regime. When making the proper correspondence to our parameter $\gamma$ and propagating the errors accordingly, they found a value of $\gamma = 0.20 \pm 1.14$. Comparing this result with our work, we observe that, while we also found a positive value for $\gamma$, our study contributed to constraining the evolution parameter, decreasing the error value.

\citet{Kett15} also does a similar study of the $M-L_X$ relation, using their newly proposed subsample of 12 intermediate-mass clusters from the CFHTLenS and XMM-CFHTLS surveys, together with 10 galaxy groups in the COSMOS field from \citet{Kettula13}, and with additional 48 high-mass clusters from CCCP \citep{Hoek15}. Notably, \citet{Kett15} represents an early study involving CODEX clusters, with X-ray luminosities estimated from XMM-Newton observations. Despite their slope $\beta$ being steeper than the self-similar prediction, it remains the most consistent with our eRASS1 result, differing by only 0.75$\sigma$. The small discrepancy probably arises from their inclusion of low-mass systems or the fact that, although they applied a Bayesian correction for the Eddington bias \citep{Vikhlinin09}, they did not specify any additional selection function or employ a statistical analysis to account for the different samples used.

We also conjecture what environmental processes in galaxy clusters could justify this deviation in slope from previous studies. Non-gravitational effects (e.g., AGN feedback) are expected to have a greater impact in low-mass clusters, such that it would be reasonable to suggest a different behavior of scaling relations in the galaxy group and galaxy cluster regimes \citep{Pratt09, Kravt12, Kett15}. For the $M-L_X$ relation, this implies a steepening in low-mass systems. As for when both groups and clusters are included, we observe a decreasing trend for the slope value \citep{Kett15}. This could support our findings that the high-mass end of the cluster population presents a relatively shallower slope.

\section{Summary and Conclusions}

This paper discusses several sets of linear fitting for calibrating the $M-L_X$ scaling relation and its evolution, represented by Eq. \ref{LxM}. We developed a Bayesian statistical model and analyzed its general impact, along with the effects of the halo mass function and selection functions on our findings. We also applied our full statistical model to combinations of our subsamples to grasp its contributions to our final result. Additionally, we performed our analysis on a subsample of clusters matched with eROSITA DR1, taking on the opportunity to improve our X-ray luminosity estimates. All the setups and results are presented in Table \ref{table_sum_all_fits}.
\begin{center}
\scriptsize
\begin{table*}[t]
\caption{Summary of all fits done in this work. The first column specifies the fitting method, whether it used a simple linear fit function, our full likelihood model, or some modification of it. More specifically, methods (i), (ii), and (iii) describe the linear fit function, the likelihood model without the selection functions, and the likelihood model disregarding only the HMF, respectively, as described in Appendix \ref{elem_stat_model}. The second column shows which cluster sample the fitting was applied to. The other columns present the best parameters found.}
\begin{tabular}{lcccccc}
\hline \hline \\[-2ex]
\multicolumn{1}{c}{Fitting model} &
\multicolumn{1}{c}{Sample} &
\multicolumn{1}{c}{$\alpha$} &
\multicolumn{1}{c}{$\beta$} &
\multicolumn{1}{c}{$\gamma$} &
\multicolumn{1}{c}{$\sigma_{intr}$}
 
\\[0.5ex] \hline
\\[-1.8ex]

Full likelihood & WL sample & $-0.12 \pm 0.02$ & $ 0.75 \pm 0.09$ & $0.65 \pm 0.43 $ & $ 0.16 \pm 0.02$ \\

Full likelihood & WL but \emph{main-subsample} & $-0.09 \pm 0.02$ & $0.75 \pm 0.11$ & $0.89 \pm 0.56$ & $0.18 \pm 0.02$ \\

Full likelihood & WL but \emph{Herbonnet20} & $-0.21 \pm 0.03$ & $0.47 \pm 0.15$ & $2.40 \pm 0.61$ & $0.12 \pm 0.02$ \\

Full likelihood & WL but \emph{Oguri21} & $-0.09 \pm 0.03$ & $0.69 \pm 0.10$ & $0.43 \pm 0.47$ & $0.16 \pm 0.02$ \\

Full likelihood & eRASS1 sample & $-0.09 \pm 0.04$ & $ 1.11 \pm 0.15$ & $0.004 \pm 0.790 $ & $ 0.17 \pm 0.04$ \\

(i) & WL sample & $-0.01 \pm 0.02$ & $0.54 \pm 0.06$ & - & - \\

(ii) & WL sample & $0.04 \pm 0.03$ & $0.30 \pm 0.05$ & $1.67 \pm 0.48$ & $0.21 \pm 0.02$ \\

(iii) & WL sample & $-0.19 \pm 0.02$ & $0.76 \pm 0.09$ & $0.14 \pm 0.47$ & $0.16 \pm 0.01$ \\

No richness PDFs & WL sample & $0.04 \pm 0.03$ & $0.28 \pm 0.05$ & $1.64 \pm 0.48$ & $0.21 \pm 0.01$ \\

Gaussian $L_X$ PDF & WL sample & $-0.13 \pm 0.02$ & $0.75 \pm 0.08$ & $0.38 \pm 0.45$ & $0.15 \pm 0.02$ \\

Fix $\beta_{\mathrm{self}}$ & WL sample & $-0.12 \pm 0.02$ & $0.9$ & $0.58 \pm 0.41$ & $0.15 \pm 0.02$ \\

$\sigma^{+}_{\lambda_{\mathrm{intr}}}$ & WL sample & $-0.03 \pm 0.02$ & $0.62 \pm 0.08$ & $1.26 \pm 0.46$ & $0.16 \pm 0.02$ \\

$\sigma^{-}_{\lambda_{\mathrm{intr}}}$ & WL sample & $-0.17 \pm 0.02$ & $0.71 \pm 0.09$ & $0.56 \pm 0.43$ & $0.18 \pm 0.02$ \\

$I_{\mathrm{H_{20}}}$ selection & WL sample & $0.09 \pm 0.02$ & $0.84 \pm 0.08$ & $0.31 \pm 0.36$ & $0.13 \pm 0.02$ \\

Full likelihood & eRASS1 but \emph{Herbonnet20} & $-0.22 \pm 0.05$ & $ 0.82 \pm 0.24$ & $2.4 \pm 1.0 $ & $ 0.14 \pm 0.04$ \\
\hline
\hline

\label{table_sum_all_fits}
\end{tabular}
\end{table*}
\end{center}

We focus our final analysis on the causes and implications of our findings for the slope $\beta$ and the evolution $\gamma$ parameters. From what was discussed and the results presented in this work, we can summarize our main discoveries as follows:

\begin{itemize}

\item When using our full Bayesian model to fit the \emph{WL sample} (main fit) we found a shallower slope than the self-similar expectation, with a 1.7$\sigma$ difference. Notably, compared to other previous works, our analysis contributed to constraining the $\gamma$ parameter, reducing the error below the parameter value. The intrinsic scatter aligns with expectations for a high redshift sample.

\item We found that including the selection functions, especially the richness PDFs, has a primary effect of steepening the scaling relation, while the HMF helps constrain the evolution parameter (further detailed in Appendix \ref{elem_stat_model}). These results suggest that the low value found for the slope in the main fit arises from the sample itself (i.e., the mass range) or unmodeled cluster properties, rather than the statistical method.

\item When analyzing the contribution of each subsample to our work, we observed that \emph{Herbonnet20} subsample has the most impact on the evolution parameter, which could be partially explained by cool core contamination. We found no major contribution of the \emph{main-subsample} for constraining $\gamma$ and found the \emph{Oguri21} subsample to have a small impact on the slope value.

\item We also performed a fit on a subsample of 42 clusters from our \emph{WL sample} that matched the eROSITA year 1 data. We take advantage of the better resolution provided by eRASS1 and compute the X-ray luminosity for this subsample. The analysis found a slope consistent with the main fit within 2$\sigma$ and aligned more closely with the self-similar prediction and previous studies. The evolution parameter also agrees with the main fit, indicating no significant evolution.
\end{itemize}

In conclusion, our main result for the $M-L_X$ calibration presents no evidence of redshift evolution and shows a shallower slope than the self-similar prediction and earlier studies. The agreements are improved when using the \emph{eRASS1 sample}, which suggests that correcting for X-ray flux point source contamination has an important effect on the slope parameter estimation. As for the difference between previous works, we note that our analysis stands out by carefully accounting for selection effects and maintaining statistical consistency between all subsamples. Even so, the observed deviation in slope may be explained by non-gravitational effects like AGN feedback, which are more significant in low-mass systems and may steepen the $M-L_X$ relation in this regime, while high-mass clusters would tend to exhibit a shallower slope.

Overall, we conclude that the $M-L_X$ calibration with high-mass galaxy clusters can result in a shallower slope value. Furthermore, different statistical approaches can significantly affect the results (e.g., the HMF and the richness uncertainties and correlation with mass). In addition to the statistical approach presented in this paper, extending the mass range of the cluster sample with reliable weak lensing mass estimations seems to be the way forward to revealing the astrophysical properties of our data.






\vspace{0.2cm}
\noindent\textit{Acknowledgments}. \footnotesize{We thank the CODEX team for creating and publishing the survey data. This work is partially based on data from eROSITA, the soft X-ray instrument aboard SRG, a joint Russian-German science mission supported by the Russian Space Agency (Roskosmos), in the interests of the Russian Academy of Sciences represented by its Space Research Institute (IKI), and the Deutsches Zentrum für Luft- und Raumfahrt (DLR). The SRG spacecraft was built by Lavochkin Association (NPOL) and its subcontractors and is operated by NPOL with support from the Max Planck Institute for Extraterrestrial Physics (MPE). The development and construction of the eROSITA X-ray instrument was led by MPE, with contributions from the Dr. Karl Remeis Observatory Bamberg \& ECAP (FAU Erlangen-Nuernberg), the University of Hamburg Observatory, the Leibniz Institute for Astrophysics Potsdam (AIP), and the Institute for Astronomy and Astrophysics of the University of Tübingen, with the support of DLR and the Max Planck Society. The Argelander Institute for Astronomy of the University of Bonn and the Ludwig Maximilians Universität Munich also participated in the science preparation for eROSITA. 
I.P. acknowledges the support from the scholarship from the Brazilian federal funding agency Coordenação de perfeiçoamento de Pessoal de Nível Superior – Brasil (CAPES). E.S.C. acknowledges the support of the funding agencies CNPq (309850/2021-5) and FAPESP (2023/02709-9). R.A.D. acknowledges support from the Conselho Nacional de Desenvolvimento Científico e Tecnológico – CNPq through BP grant 308105/2018-4, and the Financiadora de Estudos e Projetos – FINEP grants REF. 1217/13 – 01.13.0279.00 and REF 0859/10 – 01.10.0663.00. L.V.W. acknowledges the support from NSERC.}


%



\vspace{4cm}
\appendix
\section{Features of the statistical model}\label{elem_stat_model}

\normalsize We tested the impact of our general Bayesian statistical approach and, more specifically, the impact of the selection functions on the calibration of the scaling relation. To achieve this, we applied three other fits to our \emph{WL sample}: \\

(i) A simple linear fit using the Python function \textsc{lmfit.minimize}, where we disregard the redshift term in Eq. \ref{LxM} and the evolution evaluation is done separately;\\

(ii) Applying our Bayesian statistical approach but neglecting all the selection functions described in Eq. \ref{Select.}; \\

(iii) Applying our full Bayesian statistical analysis but without the Halo Mass Function (HMF).\\

For the first case, using the \textsc{lmfit.minimize}\footnote{Documentation available at \url{https://lmfit.github.io/lmfit-py/fitting.html}.} function - which ignores intrinsic scatter and mass errors - we found a slope of $0.54 \pm 0.06$, lower than our previous results and further in disagreement with self-similarity. Additionally, a linear fit of residuals versus redshift showed no strong indications of temporal evolution.

As for the second case, our model includes only the PDFs accounting for measurement errors and intrinsic scatter for $L_X$ and $M$, and the HMF. The best-fit parameters differed from the full likelihood analysis: the evolution parameter $\gamma = 1.67 \pm 0.48$ exhibited a significant increase compared to the main fit, and the slope $\beta = 0.30 \pm 0.05$ decreased by more than half the previous value.

\begin{figure}[h!]
    \centering
    \includegraphics[width=0.6\hsize]{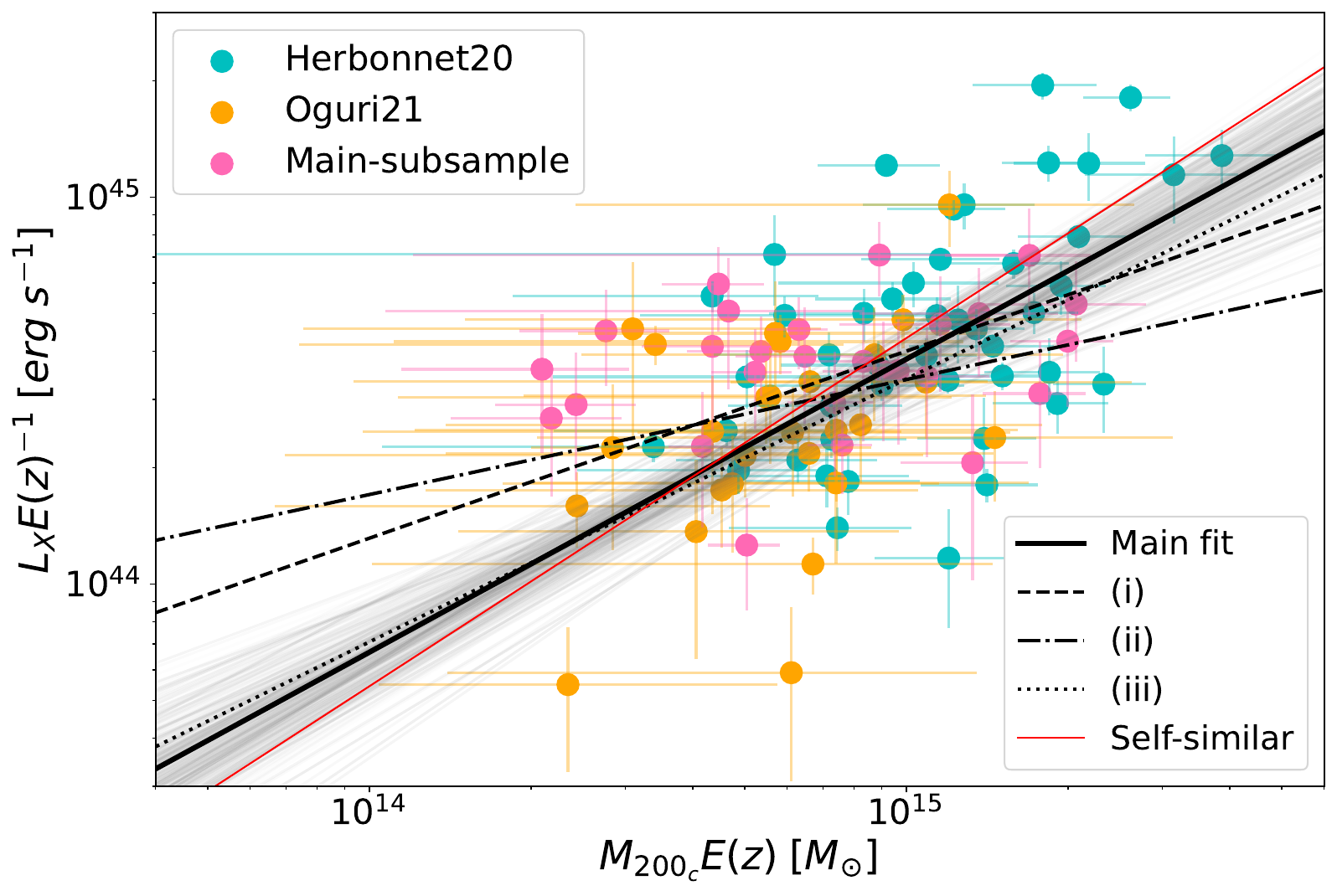}
    \caption{The \emph{WL sample}, the main fit, and the three fits discussed in this section to investigate the terms in the statistical model. The red line illustrates the slope of 0.9 predicted by self-similarity for comparison.}
    \label{4_fits}
\end{figure}

In the final approach, we excluded the halo mass function, effectively ignoring that our sample’s mass distribution is not a fair representation of the clusters in the Universe. This yielded results similar to the main fit, with $\beta = 0.76 \pm 0.09$. However, the result for the evolution parameter, $\gamma = 0.14 \pm 0.47$, is now dominated by the uncertainties. All these different fits, as well as the main fit using the full likelihood, are displayed in Fig. \ref{4_fits}, and their best parameter values are in Table \ref{table_sum_all_fits}.

These results show that including the selection functions plays a key role in steepening the $M-L_X$ scaling relation. However, the small slope value found in the main fit is unlikely due to the statistical model itself, as the simple linear fit (i) deviates from the self-similar value by approximately $6\sigma$. This suggests that our results are primarily influenced by the choice of our galaxy cluster sample, rather than the calibration method adopted.

As for the evolution parameter, all fits agree on a small positive value for $\gamma$, but the inclusion of the HMF and the selection functions contributes to constraining the parameter. In particular, a comparison between (iii) and the main fit reveals that in the former (excluding the HMF), the result is dominated by uncertainties, whereas in the latter, the error is constrained to be below the measured value.

\subsection{Impact of the richness and X-ray luminosity PDFs}\label{rich_and_lxpdf}

We have found that accounting for selection effects is crucial for constraining the slope parameter, which decreases significantly when the selection functions in Eq. \ref{Select.} are omitted. By performing another fit, we found that the main causes for such discrepancy are the richness PDFs of the observation uncertainty, $P(\Tilde{\lambda}|\lambda)$, and richness-mass relation and scatter, $P(\lambda | \langle \ln\lambda | \mu \rangle)$. This additional analysis was made by excluding only these richness PDFs and retrieved almost the same values as when we removed all the selection functions - i.e., $\beta = 0.28 \pm 0.05$ and $\gamma = 1.64 \pm 0.48$. This suggests that, at least for cluster samples built by richness-based selections, considering the richness intrinsic scatter and uncertainties in the selection functions is of great importance when calibrating the $M-L_X$ relation.

Finally, we tested modeling the X-ray luminosity PDF with a Gaussian instead of a Poisson distribution to assess potential bias in $\gamma$ (see Section \ref{Hier_model}). The results remained consistent within $1\sigma$, indicating that the choice for a Poisson distribution had a minor impact on the final calibration.

The results of the two fits discussed above are also presented in Table \ref{table_sum_all_fits}.

\subsection{The evolution parameter for a self-similar slope}

Considering the shallow slope found in our main fit, we also tested whether or not we would find any redshift evolution trend if we constrained the $\beta$ parameter to the self-similar value. There is a possibility that the $\gamma$ value was a compensation for a misestimation of the slope. With this in mind, we sampled our posterior distribution as done in the main fit, fixing the slope to the self-similar value of $0.9$. The best-fit value for each parameter is displayed in Table \ref{table_sum_all_fits}.

The $\gamma$ value found in the fit with fixed slope, $\gamma = 0.58 \pm 0.41$, agrees with the main fit under a 1$\sigma$ interval. Hence, this new result also supports the statistically significant evidence of no redshift evolution in the scaling relation.

\section{The $M-\lambda $ relation}\label{richmass}

As shown in Section \ref{results} and detailed in Appendix \ref{rich_and_lxpdf}, the correlation between cluster richness and mass plays a key role in calibrating the $M-L_X$ relation.  This motivated us to assess the impact of adopting the $M-\lambda$ relation from the literature (Equation \ref{lamb_m}). In our model, we considered a PDF for the richness given its expected value, as described in equation \ref{rich_loggauss}. Therefore, to explore how variations of the $M-\lambda$ relation impact our final results, we recomputed the main fit using the uncertainty bounds of the richness intrinsic scatter ($\sigma_{\lambda_{\mathrm{intr}}}$), as this captures deviations from the relation's parameters.

\begin{figure}[h!]
    \centering
    \includegraphics[width=0.6\hsize]{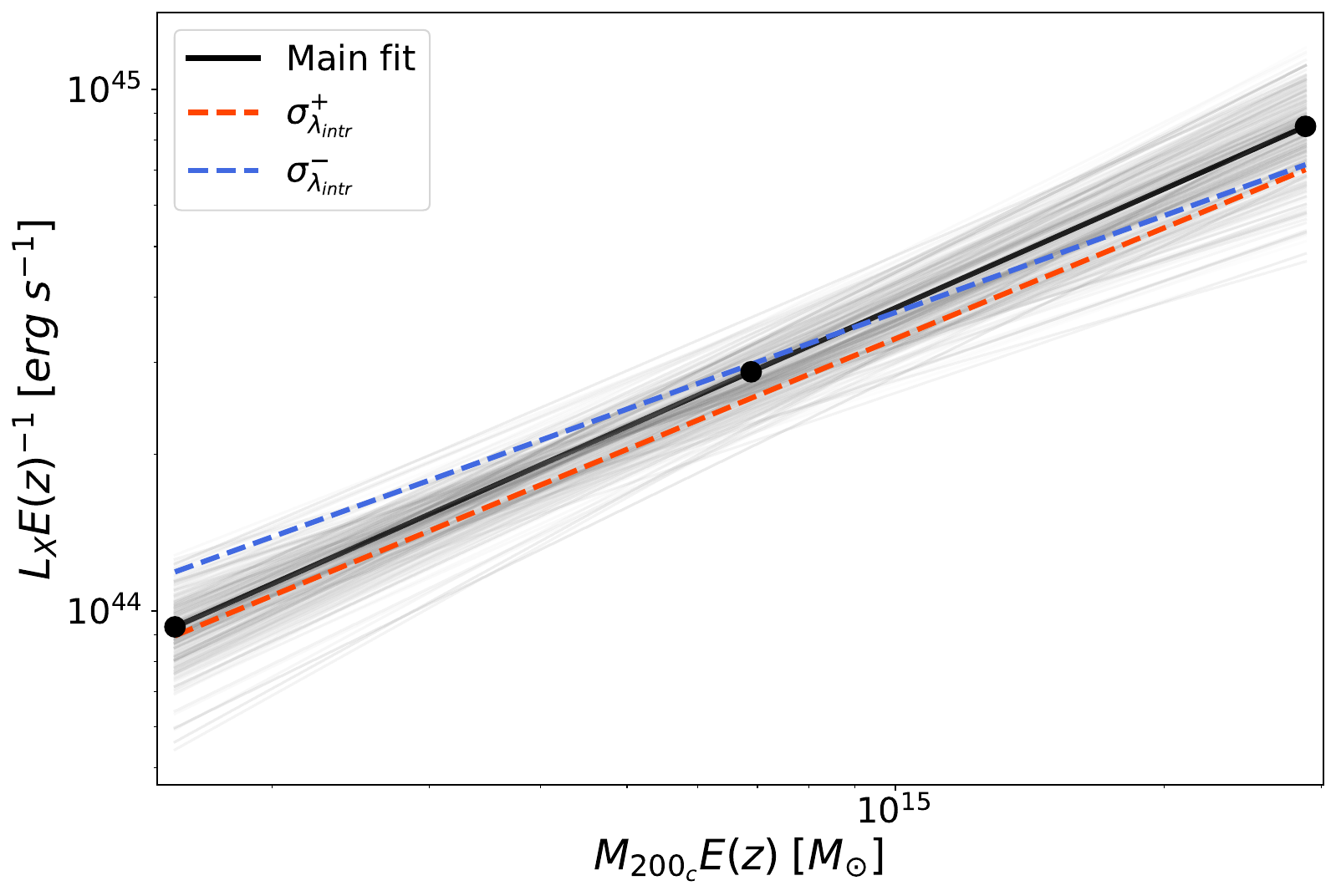}
    \caption{The main fit (\textit{black line}) and the fits applying the upper (\textit{orange dashed line}) and lower (\textit{blue dashed line}) bounds for the richness intrinsic scatter $\sigma_{\lambda_{\mathrm{intr}}}$. The results from a few of the main fit iterations are depicted by the faded gray lines, illustrating the error associated with the best fit. The three black dots represent the minimum, mean, and maximum mass values of our \textit{WL sample}.}
    \label{lamb_scatt}
\end{figure}

According to \citet{Kiiveri21}, the best-fit value for the intrinsic scatter is $\sigma_{\lambda_{\mathrm{intr}}} = 0.17^{+0.13}_{-0.09}$, setting the upper and lower bounds at $\sigma_{\lambda_{\mathrm{intr}}}^{+} = 0.30$ and $\sigma_{\lambda_{\mathrm{intr}}}^{-} = 0.09$, respectively. We than recomputed the main fit (i.e., full likelihood applied to the entire \emph{WL sample}) using these extremes, as shown in Figure \ref{lamb_scatt}. The results are displayed in Table \ref{table_sum_all_fits}. 

Compared to the main fit, we find that uncertainties in the $M$–$\lambda$ relation have no significant impact on the slope and evolution parameters. The direct comparison of the intercept values is not straightforward, as they are rescaled by pivot quantities through $\alpha_{\mathrm{scaled}} = \alpha - \log L_{\mathrm{X,piv}} + \beta \log M_{\mathrm{piv}} + \gamma \log(1+z_{\mathrm{piv}})$. However, we can compare the expected X-ray luminosity $L_{\mathrm{X,expec.}}$ given a mass value for the three fits (i.e., main fit, $\sigma_{\lambda_{\mathrm{intr}}}^{+}$, and $\sigma_{\lambda_{\mathrm{intr}}}^{-}$). We would expect a smaller deviation when considering the mean mass value of our cluster sample and a larger difference in $L_{\mathrm{X,expec.}}$ for the edges of our data points. Indeed, for the sample’s mean mass of $10^{14.84},M_{\odot}$, the maximum variation in $L_{\mathrm{X,expec.}}$ is only $0.12\%$, increasing slightly to $0.21\%$ and $0.19\%$ at the lower and upper mass ends, respectively.

By analysing the MCMC chains from the main fit — which reflects the errors associated with the best-fit parameters — we find that the variations obtained for $\sigma_{\lambda_{\mathrm{intr}}}^{+}$ and $\sigma_{\lambda_{\mathrm{intr}}}^{-}$ are well captured within 2.5$\sigma$ and 2$\sigma$ intervals for the mean mass value and at the sample boundaries, respectively. Overall, we conclude that, across the mass range considered, uncertainties in the $M–\lambda$ relation have a negligible effect on our results.

\section{The Herbonnet20 subsample}\label{H20}

As discussed in section \ref{cont_samples}, when comparing the results without \emph{Herbonnet20} with the main fit, there is a difference in the evolution parameter of 2.4$\sigma$, which could be considered a tension. To investigate this, we considered two hypotheses. First, unlike the \emph{Oguri21} subsample, the \emph{Herbonnet20} selection may be shallower than CODEX, potentially affecting our results. Since the \emph{Herbonnet20} clusters come from weak lensing follow-ups of REFLEX and NORAS at $z<0.3$, and MACS at higher redshifts, we modeled a sampling function, $P(I_{\mathrm{H_{20}}} | \Tilde{l_X}, z)$, that reflects the shallowest selection of these surveys. We also applied the appropriate corrections to ensure consistency with our $L_X$ estimates (e.g., K-corrections).

\begin{equation}
    \begin{aligned}
        P(I_{\mathrm{H_{20}}} | \Tilde{l_X}, z) = 
        \begin{cases}
            L_X > 4.5 \cdot 10^{44}\; (z/0.3)^{2.1} \; \text{for }z <0.3\\
            L_X > 3.4\cdot 10^{44}\; (z/0.3)^{2.1} \; \text{for }z >0.3\\
        \end{cases}
    \end{aligned}
\end{equation}\\

\begin{figure}[h!]
    \centering
    \includegraphics[width=0.6\hsize]{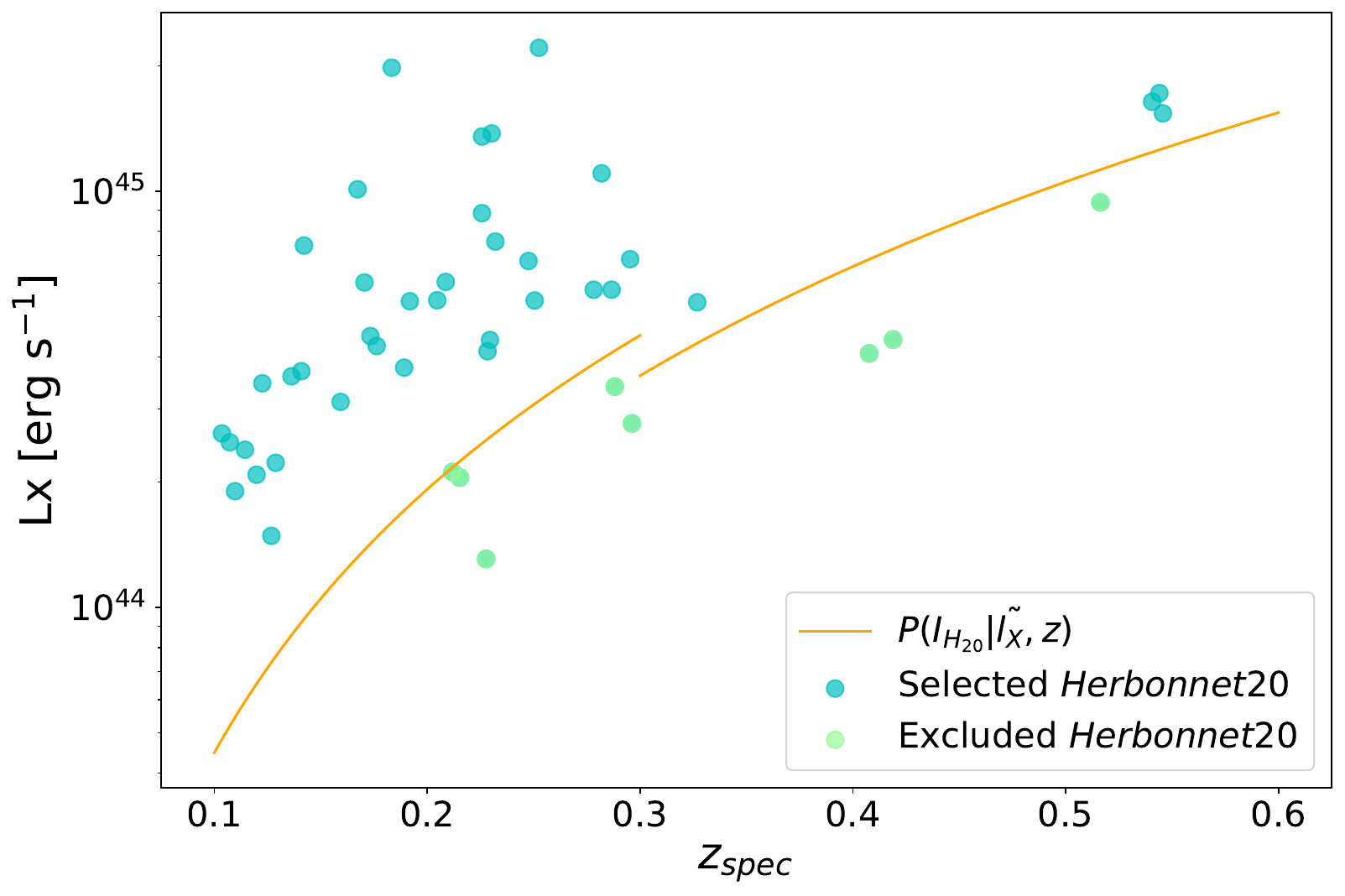}
    \caption{The \emph{Herbonnet20} subsample in the X-ray luminosity -- redshift plane. The function $P(I_{\mathrm{H_{20}}} | \Tilde{l_X}, z)$ is depicted in orange, and the colors blue and green differentiate the selected and excluded clusters, respectively.}
    \label{H20_sel}
\end{figure}

We applied this selection to the \emph{Herbonnet20} sample, which removed 8 clusters \citep[all from CCCP,][]{Hoek15}, as shown in Figure \ref{H20_sel}. We recalculated the sampling function — analogous to equation \ref{Psamp} — for the resulting subsample. Combining it with the other CODEX selection functions, we performed a fit using only the \emph{Herbonnet20} data and the \emph{WL sample} (now with 92 clusters). The results in both analyses agreed with the main fit within 1$\sigma$, indicating that the $P(I_{\mathrm{H_{20}}}| \Tilde{l_X}, z)$ selection function has minimal impact on our results. The best-fit parameters for the \emph{WL sample} are displayed in Table \ref{table_sum_all_fits}.

We also considered the possibility of the \emph{Herbonnet20} subsample being more affected by cool core contamination, which would have a lesser impact on the eRASS1 fit. Hence, we performed the fit using eROSITA luminosities for the full sample minus the \emph{Herbonnet20} subsample, and compared the results with those of the original eRASS1 fit. The results show a $\gamma$ change of 1.9$\sigma$, indicating that the tension is lessened when accounting for the flux contamination in core regions, but not entirely gone. The results of this analysis are also presented in Table \ref{table_sum_all_fits}.

\section{The CODEX-LS catalog}

In Table \ref{tab:catalogue_columns}, we describe the X-ray properties of the clean CODEX-LS Group catalog. To create the clean catalog, we applied cuts in the mask fraction (MF$<0.3$) and the redshift ($z<0.7$), and used the cluster selection described in Eq. \ref{Pclean} on the full CODEX-LS catalog. Source flux and luminosities are based on RASS data. The catalogs described in Appendix A are only available in electronic form at the CDS.

\begin{table*}[h!]

    \caption{Description of the columns of the CODEX-LS catalog.}
    \label{tab:catalogue_columns}
    \centering
    \begin{tabular*}{\textwidth}
    {l@{\extracolsep{\fill}}llr}
        \hline\hline \\[-2ex]
        \multicolumn{1}{l}{Column} & \multicolumn{1}{l}{Unit} & \multicolumn{1}{c}{Description} & \multicolumn{1}{r}{Example} \\
        \hline
        \\[-1.8ex]
        \texttt{CODEX} & & X-ray source ID in the CODEX catalog  & 93280903 \\
        \texttt{RA} & deg & X-ray detection right ascension (J2000) & 95.56809 \\
        \texttt{DEC}& deg & X-ray detection declination (J2000) & 
        --64.67731 \\
        \texttt{Z\_RED}& & Red sequence redshift & 0.0281 \\
        \texttt{ZSPEC} & & Cluster spectroscopic redshift & 0.0281 \\
        \texttt{LAMBDA\_NORM} & & Normalized cluster richness & 50. \\
        \texttt{ELAMBDA\_NORM} & & Uncertainty on Normalized cluster richness & 50. \\
        \texttt{LX0124} & ergs s$^{-1}$ & Luminosity in the (0.1-2.4) keV band of the cluster, aperture $R_\text{500c}$ & $2.3\times10^{43}$ \\
        \texttt{ELX} & ergs s$^{-1}$ & Uncertainty on \texttt{LX0124} & $7.59\times10^{41}$ \\
        \texttt{FLUX052} & ergs s$^{-1}$ cm$^{-2}$ & Galaxy cluster X-ray flux in the 0.5-2.0 keV band & $7.25\times10^{-12}$ \\
        \texttt{EFLUX052} & ergs s$^{-1}$ cm$^{-2}$ & Uncertainty on \texttt{FLUX052} & $2.39\times10^{-13}$ \\
        \texttt{LOCAL\_BKG\_FLUX} & ergs s$^{-1}$ cm$^{-2}$ & Local background flux & $0.000231$ \\
        \texttt{MASKFRAC} &  & Mask fraction & $0.015670$\\
        \hline
    \end{tabular*}
\end{table*}

\begin{table*}[h!]

    \caption{Description of the columns of the weak lensing and eROSITA data of the CODEX-LS catalog.}
    \label{tab:catalogue2_columns}
    \centering
    \begin{tabular*}{\textwidth}
    {l@{\extracolsep{\fill}}llr}
        \hline\hline \\[-2ex]
        \multicolumn{1}{l}{Column} & \multicolumn{1}{l}{Unit} & \multicolumn{1}{c}{Description} & \multicolumn{1}{r}{Example} \\
        \hline 
        \\[-1.8ex]
        \texttt{CODEX} & & X-ray source ID in the CODEX catalog  & 93280903 \\
        \texttt{RA} & deg & X-ray detection right ascension in eRASS1 (J2000) & 95.56809 \\
        \texttt{DEC} & deg & X-ray detection declination in eRASS1 (J2000) & 
        --64.67731 \\
        \texttt{M\_WL} & $M_\odot$ & Weak lensing cluster mass encompassing an overdensity of 200 times the critical & 1.e15 \\
        \texttt{M\_WL\_ERRP} & $M_\odot$ & Error $+$ on \texttt{M\_WL}  & 1.e14 \\
        \texttt{M\_WL\_ERRM} & $M_\odot$ & Error $-$ on \texttt{M\_WL}  & 1.e14 \\
        \texttt{WL\_SUBSAMPLE} &  & Source of \texttt{M\_WL}; Herbonnet20 (H), Oguri21 (O), Main-subsample (Ms) & Ms \\
        \texttt{LX0124} & ergs s$^{-1}$ & eRASS1 Luminosity in the (0.1-2.4) keV band of the cluster, aperture $R_\text{500c}$ & $2.3\times10^{43}$ \\
        \texttt{ELX} & ergs s$^{-1}$ & Uncertainty on \texttt{LX0124} & $7.59\times10^{41}$ \\
        \texttt{FLUX0623} & ergs s$^{-1}$ cm$^{-2}$ & Galaxy cluster X-ray flux in the 0.6-2.3 keV band of eRASS1 & $7.25\times10^{-12}$ \\
        \texttt{EFLUX0623} & ergs s$^{-1}$ cm$^{-2}$ & Uncertainty on \texttt{FLUX0623} & $2.39\times10^{-13}$ \\
        \hline
    \end{tabular*}
\end{table*}

\vspace{4cm}
\bibliography{sample631}{}

\begin{thebibliography}{}
\expandafter\ifx\csname natexlab\endcsname\relax\def\natexlab#1{#1}\fi
\providecommand{\url}[1]{\href{#1}{#1}}
\providecommand{\dodoi}[1]{doi:~\href{http://doi.org/#1}{\nolinkurl{#1}}}
\providecommand{\doeprint}[1]{\href{http://ascl.net/#1}{\nolinkurl{http://ascl.net/#1}}}
\providecommand{\doarXiv}[1]{\href{https://arxiv.org/abs/#1}{\nolinkurl{https://arxiv.org/abs/#1}}}

\bibitem[{{Allen} {et~al.}(2011){Allen}, {Evrard}, \& {Mantz}}]{Allen11}
{Allen}, S.~W., {Evrard}, A.~E., \& {Mantz}, A.~B. 2011, \araa, 49, 409,
  \dodoi{10.1146/annurev-astro-081710-102514}

\bibitem[{{Ascasibar} {et~al.}(2006){Ascasibar}, {Sevilla}, {Yepes},
  {M{\"u}ller}, \& {Gottl{\"o}ber}}]{Ascasibar06}
{Ascasibar}, Y., {Sevilla}, R., {Yepes}, G., {M{\"u}ller}, V., \&
  {Gottl{\"o}ber}, S. 2006, \mnras, 371, 193,
  \dodoi{10.1111/j.1365-2966.2006.10596.x}

\bibitem[{{Blumenthal} {et~al.}(1984){Blumenthal}, {Faber}, {Primack}, \&
  {Rees}}]{Blumenthal84}
{Blumenthal}, G.~R., {Faber}, S.~M., {Primack}, J.~R., \& {Rees}, M.~J. 1984,
  \nat, 311, 517, \dodoi{10.1038/311517a0}

\bibitem[{{Braspenning} {et~al.}(2023){Braspenning}, {Schaye}, {Schaller},
  {McCarthy}, {Kay}, {Helly}, {Kugel}, {Elbers}, {Frenk}, {Kwan}, {Salcido},
  {van Daalen}, \& {Vandenbroucke}}]{Braspenning23}
{Braspenning}, J., {Schaye}, J., {Schaller}, M., {et~al.} 2023, arXiv e-prints,
  arXiv:2312.08277, \dodoi{10.48550/arXiv.2312.08277}

\bibitem[{{Bulbul} {et~al.}(2019){Bulbul}, {Chiu}, {Mohr}, {McDonald},
  {Benson}, {Bautz}, {Bayliss}, {Bleem}, {Brodwin}, {Bocquet}, {Capasso},
  {Dietrich}, {Forman}, {Hlavacek-Larrondo}, {Holzapfel}, {Khullar}, {Klein},
  {Kraft}, {Miller}, {Reichardt}, {Saro}, {Sharon}, {Stalder}, {Schrabback}, \&
  {Stanford}}]{Bulbul19}
{Bulbul}, E., {Chiu}, I.~N., {Mohr}, J.~J., {et~al.} 2019, \apj, 871, 50,
  \dodoi{10.3847/1538-4357/aaf230}

\bibitem[{{Chen} {et~al.}(2007){Chen}, {Reiprich}, {B{\"o}hringer}, {Ikebe}, \&
  {Zhang}}]{Chen07}
{Chen}, Y., {Reiprich}, T.~H., {B{\"o}hringer}, H., {Ikebe}, Y., \& {Zhang},
  Y.~Y. 2007, \aap, 466, 805, \dodoi{10.1051/0004-6361:20066471}

\bibitem[{{Cibirka} {et~al.}(2017){Cibirka}, {Cypriano}, {Brimioulle}, {Gruen},
  {Erben}, {van Waerbeke}, {Miller}, {Finoguenov}, {Kirkpatrick}, {Henry},
  {Rykoff}, {Rozo}, {Dupke}, {Kneib}, {Shan}, \& {Spinelli}}]{Cibirka17}
{Cibirka}, N., {Cypriano}, E.~S., {Brimioulle}, F., {et~al.} 2017, \mnras, 468,
  1092, \dodoi{10.1093/mnras/stx484}

\bibitem[{{Clerc} \& {Finoguenov}(2023)}]{CF}
{Clerc}, N., \& {Finoguenov}, A. 2023, in Handbook of X-ray and Gamma-ray
  Astrophysics, 123, \dodoi{10.1007/978-981-16-4544-0_117-1}

\bibitem[{{Damsted} {et~al.}(2023){Damsted}, {Finoguenov}, {Clerc},
  {Davalgait{\.{e}}}, {Kirkpatrick}, {Mamon}, {Ider Chitham}, {Kiiveri},
  {Comparat}, \& {Collins}}]{Damsted23}
{Damsted}, S., {Finoguenov}, A., {Clerc}, N., {et~al.} 2023, \aap, 676, A127,
  \dodoi{10.1051/0004-6361/202245308}

\bibitem[{{Damsted} {et~al.}(2024){Damsted}, {Finoguenov}, {Lietzen}, {Mamon},
  {Comparat}, {Tempel}, {Dmitrieva}, {Clerc}, {Collins}, {Gozaliasl}, \&
  {Eckert}}]{Damsted24}
{Damsted}, S., {Finoguenov}, A., {Lietzen}, H., {et~al.} 2024, \aap, 690, A52,
  \dodoi{10.1051/0004-6361/202449591}

\bibitem[{Dawson(2019)}]{Dawson19}
Dawson, C. 2019, Introduction to Research Methods 5th Edition: A Practical
  Guide for Anyone Undertaking a Research Project (Little, Brown Book Group).
\newblock \url{https://books.google.com.br/books?id=pu5MDwAAQBAJ}

\bibitem[{Dey {et~al.}(2019)Dey, Schlegel, Lang, Blum, Burleigh, Fan, Findlay,
  Finkbeiner, Herrera, Juneau, {et~al.}}]{dey2019overview}
Dey, A., Schlegel, D.~J., Lang, D., {et~al.} 2019, The Astronomical Journal,
  157, 168

\bibitem[{{Eckmiller} {et~al.}(2011){Eckmiller}, {Hudson}, \&
  {Reiprich}}]{Eckmiller11}
{Eckmiller}, H.~J., {Hudson}, D.~S., \& {Reiprich}, T.~H. 2011, \aap, 535,
  A105, \dodoi{10.1051/0004-6361/201116734}

\bibitem[{{Finoguenov} {et~al.}(2020){Finoguenov}, {Rykoff}, {Clerc},
  {Costanzi}, {Hagstotz}, {Ider Chitham}, {Kiiveri}, {Kirkpatrick}, {Capasso},
  {Comparat}, {Damsted}, {Dupke}, {Erfanianfar}, {Patrick Henry}, {Kaefer},
  {Kneib}, {Lindholm}, {Rozo}, {van Waerbeke}, \& {Weller}}]{Finoguenov2020}
{Finoguenov}, A., {Rykoff}, E., {Clerc}, N., {et~al.} 2020, \aap, 638, A114,
  \dodoi{10.1051/0004-6361/201937283}

\bibitem[{{Foreman-Mackey} {et~al.}(2013){Foreman-Mackey}, {Hogg}, {Lang}, \&
  {Goodman}}]{ForMac13}
{Foreman-Mackey}, D., {Hogg}, D.~W., {Lang}, D., \& {Goodman}, J. 2013, \pasp,
  125, 306, \dodoi{10.1086/670067}

\bibitem[{Fujita \& Aung(2019)}]{Fujita19}
Fujita, Y., \& Aung, H. 2019, The Astrophysical Journal, 875, 26,
  \dodoi{10.3847/1538-4357/ab0e02}

\bibitem[{{Goodman} \& {Weare}(2010)}]{GoodmanWeare10}
{Goodman}, J., \& {Weare}, J. 2010, Communications in Applied Mathematics and
  Computational Science, 5, 65, \dodoi{10.2140/camcos.2010.5.65}

\bibitem[{Gregory(2005)}]{Gregory}
Gregory, P. 2005, Bayesian Logical Data Analysis for the Physical Sciences
  (Cambridge University Press)

\bibitem[{{Gruen} {et~al.}(2015){Gruen}, {Seitz}, {Becker}, {Friedrich}, \&
  {Mana}}]{Gruen15}
{Gruen}, D., {Seitz}, S., {Becker}, M.~R., {Friedrich}, O., \& {Mana}, A. 2015,
  \mnras, 449, 4264, \dodoi{10.1093/mnras/stv532}

\bibitem[{{Herbonnet} {et~al.}(2020){Herbonnet}, {Sif{\'o}n}, {Hoekstra},
  {Bah{\'e}}, {van der Burg}, {Melin}, {von der Linden}, {Sand}, {Kay}, \&
  {Barnes}}]{Herbonnet20}
{Herbonnet}, R., {Sif{\'o}n}, C., {Hoekstra}, H., {et~al.} 2020, \mnras, 497,
  4684, \dodoi{10.1093/mnras/staa2303}

\bibitem[{{Hoekstra} {et~al.}(2015){Hoekstra}, {Herbonnet}, {Muzzin}, {Babul},
  {Mahdavi}, {Viola}, \& {Cacciato}}]{Hoek15}
{Hoekstra}, H., {Herbonnet}, R., {Muzzin}, A., {et~al.} 2015, \mnras, 449, 685,
  \dodoi{10.1093/mnras/stv275}

\bibitem[{Ider~Chitham {et~al.}(2020)Ider~Chitham, Comparat, Finoguenov, Clerc,
  Kirkpatrick, Damsted, Kukkola, Nandra, Merloni,
  {et~al.}}]{ider2020cosmological}
Ider~Chitham, J., Comparat, J., Finoguenov, A., {et~al.} 2020, Monthly Notices
  of the Royal Astronomical Society, 499, 4768

\bibitem[{Jones {et~al.}(1998)Jones, Scharf, Ebeling, Perlman, Wegner, Malkan,
  \& Horner}]{Jones_1998}
Jones, L.~R., Scharf, C., Ebeling, H., {et~al.} 1998, The Astrophysical
  Journal, 495, 100, \dodoi{10.1086/305283}

\bibitem[{{Kaiser}(1986)}]{Kaiser86}
{Kaiser}, N. 1986, \mnras, 222, 323, \dodoi{10.1093/mnras/222.2.323}

\bibitem[{{Kelly}(2007)}]{Kelly07}
{Kelly}, B.~C. 2007, \apj, 665, 1489, \dodoi{10.1086/519947}

\bibitem[{{Kettula} {et~al.}(2013){Kettula}, {Finoguenov}, {Massey}, {Rhodes},
  {Hoekstra}, {Taylor}, {Spinelli}, {Tanaka}, {Ilbert}, {Capak}, {McCracken},
  \& {Koekemoer}}]{Kettula13}
{Kettula}, K., {Finoguenov}, A., {Massey}, R., {et~al.} 2013, \apj, 778, 74,
  \dodoi{10.1088/0004-637X/778/1/74}

\bibitem[{{Kettula} {et~al.}(2015){Kettula}, {Giodini}, {van Uitert},
  {Hoekstra}, {Finoguenov}, {Lerchster}, {Erben}, {Heymans}, {Hildebrandt},
  {Kitching}, {Mahdavi}, {Mellier}, {Miller}, {Mirkazemi}, {Van Waerbeke},
  {Coupon}, {Egami}, {Fu}, {Hudson}, {Kneib}, {Kuijken}, {McCracken},
  {Pereira}, {Rowe}, {Schrabback}, {Tanaka}, \& {Velander}}]{Kett15}
{Kettula}, K., {Giodini}, S., {van Uitert}, E., {et~al.} 2015, \mnras, 451,
  1460, \dodoi{10.1093/mnras/stv923}

\bibitem[{{Khalil} {et~al.}(2024){Khalil}, {Finoguenov}, {Tempel}, \&
  {Mamon}}]{Khalil24}
{Khalil}, H., {Finoguenov}, A., {Tempel}, E., \& {Mamon}, G.~A. 2024, \aap,
  690, A212, \dodoi{10.1051/0004-6361/202450060}

\bibitem[{{Kiiveri} {et~al.}(2021){Kiiveri}, {Gruen}, {Finoguenov}, {Erben},
  {van Waerbeke}, {Rykoff}, {Miller}, {Hagstotz}, {Dupke}, {Patrick Henry},
  {Kneib}, {Gozaliasl}, {Kirkpatrick}, {Cibirka}, {Clerc}, {Costanzi},
  {Cypriano}, {Rozo}, {Shan}, {Spinelli}, {Valiviita}, \& {Weller}}]{Kiiveri21}
{Kiiveri}, K., {Gruen}, D., {Finoguenov}, A., {et~al.} 2021, \mnras, 502, 1494,
  \dodoi{10.1093/mnras/staa3936}

\bibitem[{{Klein} {et~al.}(2019){Klein}, {Grandis}, {Mohr}, {Paulus}, {Abbott},
  {Annis}, {Avila}, {Bertin}, {Brooks}, {Buckley-Geer}, {Rosell}, {Kind},
  {Carretero}, {Castander}, {Cunha}, {D'Andrea}, {da Costa}, {De Vicente},
  {Desai}, {Diehl}, {Dietrich}, {Doel}, {Evrard}, {Flaugher}, {Fosalba},
  {Frieman}, {Garc{\'\i}a-Bellido}, {Gaztanaga}, {Giles}, {Gruen}, {Gruendl},
  {Gschwend}, {Gutierrez}, {Hartley}, {Hollowood}, {Honscheid}, {Hoyle},
  {James}, {Jeltema}, {Kuehn}, {Kuropatkin}, {Lima}, {Maia}, {March},
  {Marshall}, {Menanteau}, {Miquel}, {Ogando}, {Plazas}, {Romer}, {Roodman},
  {Sanchez}, {Scarpine}, {Schindler}, {Serrano}, {Sevilla-Noarbe}, {Smith},
  {Smith}, {Soares-Santos}, {Sobreira}, {Suchyta}, {Swanson}, {Tarle},
  {Thomas}, {Vikram}, \& {DES Collaboration}}]{Klein19}
{Klein}, M., {Grandis}, S., {Mohr}, J.~J., {et~al.} 2019, \mnras, 488, 739,
  \dodoi{10.1093/mnras/stz1463}

\bibitem[{{Kluge} {et~al.}(2024){Kluge}, {Comparat}, {Liu}, {Balzer}, {Bulbul},
  {Ider Chitham}, {Ghirardini}, {Garrel}, {Bahar}, {Artis}, {Bender}, {Clerc},
  {Dwelly}, {Fabricius}, {Grandis}, {Hern{\'a}ndez-Lang}, {Hill}, {Joshi},
  {Lamer}, {Merloni}, {Nandra}, {Pacaud}, {Predehl}, {Ramos-Ceja}, {Reiprich},
  {Salvato}, {Sanders}, {Schrabback}, {Seppi}, {Zelmer}, {Zenteno}, \&
  {Zhang}}]{Kluge2024}
{Kluge}, M., {Comparat}, J., {Liu}, A., {et~al.} 2024, \aap, 688, A210,
  \dodoi{10.1051/0004-6361/202349031}

\bibitem[{{Kravtsov} \& {Borgani}(2012)}]{Kravt12}
{Kravtsov}, A.~V., \& {Borgani}, S. 2012, \araa, 50, 353,
  \dodoi{10.1146/annurev-astro-081811-125502}

\bibitem[{{Leauthaud} {et~al.}(2010){Leauthaud}, {Finoguenov}, {Kneib},
  {Taylor}, {Massey}, {Rhodes}, {Ilbert}, {Bundy}, {Tinker}, {George}, {Capak},
  {Koekemoer}, {Johnston}, {Zhang}, {Cappelluti}, {Ellis}, {Elvis}, {Giodini},
  {Heymans}, {Le F{\`e}vre}, {Lilly}, {McCracken}, {Mellier},
  {R{\'e}fr{\'e}gier}, {Salvato}, {Scoville}, {Smoot}, {Tanaka}, {Van
  Waerbeke}, \& {Wolk}}]{Lea10}
{Leauthaud}, A., {Finoguenov}, A., {Kneib}, J.-P., {et~al.} 2010, \apj, 709,
  97, \dodoi{10.1088/0004-637X/709/1/97}

\bibitem[{{Lovisari} {et~al.}(2021){Lovisari}, {Ettori}, {Gaspari}, \&
  {Giles}}]{Lovisari21}
{Lovisari}, L., {Ettori}, S., {Gaspari}, M., \& {Giles}, P.~A. 2021, Universe,
  7, 139, \dodoi{10.3390/universe7050139}

\bibitem[{{Lovisari} {et~al.}(2020){Lovisari}, {Schellenberger}, {Sereno},
  {Ettori}, {Pratt}, {Forman}, {Jones}, {Andrade-Santos}, {Randall}, \&
  {Kraft}}]{Lovisari20}
{Lovisari}, L., {Schellenberger}, G., {Sereno}, M., {et~al.} 2020, \apj, 892,
  102, \dodoi{10.3847/1538-4357/ab7997}

\bibitem[{{Merloni} {et~al.}(2024){Merloni}, {Lamer}, {Liu}, {Ramos-Ceja},
  {Brunner}, {Bulbul}, {Dennerl}, {Doroshenko}, {Freyberg}, {Friedrich},
  {Gatuzz}, {Georgakakis}, {Haberl}, {Igo}, {Kreykenbohm}, {Liu}, {Maitra},
  {Malyali}, {Mayer}, {Nandra}, {Predehl}, {Robrade}, {Salvato}, {Sanders},
  {Stewart}, {Tub{\'\i}n-Arenas}, {Weber}, {Wilms}, {Arcodia}, {Artis},
  {Aschersleben}, {Avakyan}, {Aydar}, {Bahar}, {Balzer}, {Becker}, {Berger},
  {Boller}, {Bornemann}, {Br{\"u}ggen}, {Brusa}, {Buchner}, {Burwitz},
  {Camilloni}, {Clerc}, {Comparat}, {Coutinho}, {Czesla}, {Dannhauer},
  {Dauner}, {Dauser}, {Dietl}, {Dolag}, {Dwelly}, {Egg}, {Ehl}, {Freund},
  {Friedrich}, {Gaida}, {Garrel}, {Ghirardini}, {Gokus}, {Gr{\"u}nwald},
  {Grandis}, {Grotova}, {Gruen}, {Gueguen}, {H{\"a}mmerich}, {Hamaus},
  {Hasinger}, {Haubner}, {Homan}, {Ider Chitham}, {Joseph}, {Joyce},
  {K{\"o}nig}, {Kaltenbrunner}, {Khokhriakova}, {Kink}, {Kirsch}, {Kluge},
  {Knies}, {Krippendorf}, {Krumpe}, {Kurpas}, {Li}, {Liu}, {Locatelli},
  {Lorenz}, {M{\"u}ller}, {Magaudda}, {Mannes}, {McCall}, {Meidinger},
  {Michailidis}, {Migkas}, {Mu{\~n}oz-Giraldo}, {Musiimenta}, {Nguyen-Dang},
  {Ni}, {Olechowska}, {Ota}, {Pacaud}, {Pasini}, {Perinati}, {Pires},
  {Pommranz}, {Ponti}, {Poppenhaeger}, {P{\"u}hlhofer}, {Rau}, {Reh},
  {Reiprich}, {Roster}, {Saeedi}, {Santangelo}, {Sasaki}, {Schmitt},
  {Schneider}, {Schrabback}, {Schuster}, {Schwope}, {Seppi}, {Serim},
  {Shreeram}, {Sokolova-Lapa}, {Starck}, {Stelzer}, {Stierhof}, {Suleimanov},
  {Tenzer}, {Traulsen}, {Tr{\"u}mper}, {Tsuge}, {Urrutia}, {Veronica},
  {Waddell}, {Willer}, {Wolf}, {Yeung}, {Zainab}, {Zangrandi}, {Zhang},
  {Zhang}, \& {Zheng}}]{eRASS1}
{Merloni}, A., {Lamer}, G., {Liu}, T., {et~al.} 2024, \aap, 682, A34,
  \dodoi{10.1051/0004-6361/202347165}

\bibitem[{{Nagai}(2006)}]{Nagai06}
{Nagai}, D. 2006, \apj, 650, 538, \dodoi{10.1086/506467}

\bibitem[{{Narayan} \& {Bartelmann}(1996)}]{NARAYAN95}
{Narayan}, R., \& {Bartelmann}, M. 1996, arXiv e-prints, astro,
  \dodoi{10.48550/arXiv.astro-ph/9606001}

\bibitem[{{Navarro} {et~al.}(1997){Navarro}, {Frenk}, \& {White}}]{NFW97}
{Navarro}, J.~F., {Frenk}, C.~S., \& {White}, S. D.~M. 1997, \apj, 490, 493,
  \dodoi{10.1086/304888}

\bibitem[{{Oguri} {et~al.}(2021){Oguri}, {Miyazaki}, {Li}, {Luo}, {Mitsuishi},
  {Miyatake}, {More}, {Nishizawa}, {Okabe}, {Ota}, {Plazas Malag{\'o}n}, \&
  {Utsumi}}]{Oguri21}
{Oguri}, M., {Miyazaki}, S., {Li}, X., {et~al.} 2021, \pasj, 73, 817,
  \dodoi{10.1093/pasj/psab047}

\bibitem[{{Planelles} {et~al.}(2015){Planelles}, {Schleicher}, \&
  {Bykov}}]{Planelles15}
{Planelles}, S., {Schleicher}, D.~R.~G., \& {Bykov}, A.~M. 2015, \ssr, 188, 93,
  \dodoi{10.1007/s11214-014-0045-7}

\bibitem[{{Pratt} {et~al.}(2009){Pratt}, {Croston}, {Arnaud}, \&
  {B{\"o}hringer}}]{Pratt09}
{Pratt}, G.~W., {Croston}, J.~H., {Arnaud}, M., \& {B{\"o}hringer}, H. 2009,
  \aap, 498, 361, \dodoi{10.1051/0004-6361/200810994}

\bibitem[{{Press} \& {Schechter}(1974)}]{PS74}
{Press}, W.~H., \& {Schechter}, P. 1974, \apj, 187, 425, \dodoi{10.1086/152650}

\bibitem[{{Rykoff} {et~al.}(2014){Rykoff}, {Rozo}, {Busha}, {Cunha},
  {Finoguenov}, {Evrard}, {Hao}, {Koester}, {Leauthaud}, {Nord}, {Pierre},
  {Reddick}, {Sadibekova}, {Sheldon}, \& {Wechsler}}]{Ry14}
{Rykoff}, E.~S., {Rozo}, E., {Busha}, M.~T., {et~al.} 2014, \apj, 785, 104,
  \dodoi{10.1088/0004-637X/785/2/104}

\bibitem[{{Sereno}(2016)}]{Sereno16}
{Sereno}, M. 2016, \mnras, 455, 2149, \dodoi{10.1093/mnras/stv2374}

\bibitem[{{Tinker} {et~al.}(2008){Tinker}, {Kravtsov}, {Klypin}, {Abazajian},
  {Warren}, {Yepes}, {Gottl{\"o}ber}, \& {Holz}}]{Tinker08}
{Tinker}, J., {Kravtsov}, A.~V., {Klypin}, A., {et~al.} 2008, \apj, 688, 709,
  \dodoi{10.1086/591439}

\bibitem[{{Vikhlinin} {et~al.}(1998){Vikhlinin}, {McNamara}, {Forman}, {Jones},
  {Quintana}, \& {Hornstrup}}]{Vikh98}
{Vikhlinin}, A., {McNamara}, B.~R., {Forman}, W., {et~al.} 1998, \apj, 502,
  558, \dodoi{10.1086/305951}

\bibitem[{{Vikhlinin} {et~al.}(2009){Vikhlinin}, {Burenin}, {Ebeling},
  {Forman}, {Hornstrup}, {Jones}, {Kravtsov}, {Murray}, {Nagai}, {Quintana}, \&
  {Voevodkin}}]{Vikhlinin09}
{Vikhlinin}, A., {Burenin}, R.~A., {Ebeling}, H., {et~al.} 2009, \apj, 692,
  1033, \dodoi{10.1088/0004-637X/692/2/1033}

\bibitem[{{Voges} {et~al.}(1999){Voges}, {Aschenbach}, {Boller},
  {Br{\"a}uninger}, {Briel}, {Burkert}, {Dennerl}, {Englhauser}, {Gruber},
  {Haberl}, {Hartner}, {Hasinger}, {K{\"u}rster}, {Pfeffermann}, {Pietsch},
  {Predehl}, {Rosso}, {Schmitt}, {Tr{\"u}mper}, \& {Zimmermann}}]{Voges99}
{Voges}, W., {Aschenbach}, B., {Boller}, T., {et~al.} 1999, \aap, 349, 389,
  \dodoi{10.48550/arXiv.astro-ph/9909315}

\bibitem[{{White} \& {Rees}(1978)}]{White&Rees78}
{White}, S.~D.~M., \& {Rees}, M.~J. 1978, \mnras, 183, 341,
  \dodoi{10.1093/mnras/183.3.341}

\bibitem[{Witte \& Witte(2017)}]{Witte17}
Witte, R., \& Witte, J. 2017, Statistics (Wiley).
\newblock \url{https://books.google.com.br/books?id=KcxjDwAAQBAJ}

\end{thebibliography}
\bibliographystyle{aasjournal}



\end{document}